\documentclass[acmsmall,nonacm]{acmart}

\usepackage{natbib}
\usepackage{booktabs} 
\usepackage{tikz}
\usepackage{graphicx}
\usepackage{bm}
\usepackage[export]{adjustbox}
\usepackage{color}
\usepackage{comment}

	\let\labelindent\relax
\usepackage{enumitem}
\usepackage[normalem]{ulem}
\usepackage{pbox}
\usepackage{multirow}
\usepackage{makecell}
\usepackage[caption=false]{subfig}
\usepackage{mathtools}

\usepackage{algpseudocode}
\usepackage{color}
\usepackage{array}

\usepackage{subfloat}
\usepackage{amsfonts}

\usepackage[vlined,ruled]{algorithm2e}

\graphicspath{results/plots}

\algblock{ParFor}{EndParFor}
\algnewcommand\algorithmicparfor{\textbf{parallel for}}
\algnewcommand\algorithmicpardo{\textbf{do}}
\algnewcommand\algorithmicendparfor{\textbf{end\ parallel for}}
\algrenewtext{ParFor}[1]{\algorithmicparfor\ #1\ \algorithmicpardo}
\algrenewtext{EndParFor}{\algorithmicendparfor}

\newcolumntype{L}[1]{>{\raggedright\let\newline\\\arraybackslash\hspace{0pt}}m{#1}}
\newcolumntype{C}[1]{>{\centering\let\newline\\\arraybackslash\hspace{0pt}}m{#1}}
\newcolumntype{R}[1]{>{\raggedleft\let\newline\\\arraybackslash\hspace{0pt}}m{#1}}
\usepackage{boxhandler}

\begin{document}

\title{ButterFly BFS - An Efficient Communication Pattern for Multi Node Traversals}


\author{Oded Green}
\affiliation{%
	  \institution{NVIDIA (ogreen@nvidia.com ) and  
      Computational Science and Engineering, Georgia Institute of Technology}
}



\begin{abstract}

    Breadth-First Search (BFS) is a building block used in a wide array of graph analytics and is used in various network analysis domains: social, road, transportation, communication, and much more.
    Over the last two decades, network sizes have continued to grow.  The popularity of BFS has brought with it a need for significantly faster traversals.
    Thus, BFS algorithms have been designed to exploit shared-memory and shared-nothing systems--this includes algorithms for accelerators such as the GPU. GPUs offer extremely fast traversals at the cost of processing smaller graphs due to their limited memory size. In contrast, CPU shared-memory systems can scale to graphs with several billion edges but do not have enough compute resources needed for fast traversals. This paper introduces ButterFly BFS, a multi-GPU traversal algorithm that allows analyzing significantly larger networks at high rates. ButterFly BFS scales to the similar-sized graphs processed by shared-memory systems while improving performance by more than 10X compared to CPUs.
    We evaluate our new algorithm on an NVIDIA DGX-2 server with 16 V100 GPUS and show that our algorithm scales with an increase in the number of GPUS. We show that we can achieve a roughly $70\%$ performance linear speedup, which is non-trivial for BFS. For a scale 29 Kronecker graph and edge factor of 8, our new algorithm traverses the graph at a rate of over 300 GTEP/s. That is a high traversal rate for a single server.
    
    \end{abstract}

\maketitle

\section{Introduction}
\label{sec:intro}

Graph algorithms have been developed now for well over fifty years.
Graphs are used in a wide range of applications to represent various datasets as they enable simplifying problem statements. For example, finding the number of people connected two or three hops away from a person of interest might require multiple queries if stated as a SQL problem. In contrast, graphs allow stating the problem differently.  A breadth-first search (also commonly referred to as a BFS traversal) finds the distances between the players within a network. BFS is a key building block for a wide range of applications, including all-pair shortest-paths \cite{Warshall1962}, betweenness centrality (static \cite{Brandes2001}  and dynamic \cite{GreenStreaming}), finding both weakly \cite{slota2014bfs} and strongly \cite{fleischer2000identifying} connected components, $s-t$ connectivity \cite{bader2006designing}, and much more. Numerous formulations for BFS exist, including vertex-centric, gather-apply-scatter (GAS), or algebra operations (often referred to as BLAS).

As graph sizes continue to grow and the need for close to real-time results are needed, optimizing the BFS traversal is especially important. Algorithms have been designed to exploit massively multi-threaded shared memory systems as the Cray XMT \cite{bader2006designing}, NVIDIA GPU's \cite{MerrillGPU}, and the EMU Chick \cite{hein2018programming}. While many algorithms focus on the classical top-down BFS traversals, the introduction of the direction optimizing traversal first introduced by Beamer \emph{et al.} \cite{beamer2012direction} has brought a plethora of BFS implementations. Direction optimizing BFS accelerates the speed of the BFS by avoiding traversing unnecessary edges. Therefore, it is not surprising that many Graph500 implementations use the direction optimizing BFS traversal. At the same time, direction optimizing BFS traversal is not applicable for a wide range of all-pairs shortest-path problems that require traversing all the edges.


Other algorithms have focused on designing BFS to work on distributed systems \cite{Buluc01112011,bulucc2011parallel,beamer2013distributed} (often referred to as shared-nothing system as the servers can only communicate via an interconnect).
Distributed  BFS implementations suffer from a common performance side-effect. Namely, the performance drops significantly (by as much as an order of magnitude) when the execution moves from a single compute node to multiple compute nodes. The performance drop is in part due to communication overheads such as network bandwidth and latency.

\subsection*{Algorithmic Contributions}

1.We introduce Butterfly BFS, a novel approach for scaling BFS to multiple compute nodes. Butterfly BFS uses a butterfly network for synchronizing the active vertex frontier across the multiple compute nodes. The butterfly network replaces the all-to-all communication pattern and reduces the number of messages sent and the amount of transmitted data. 

2. We show that the butterfly communication pattern supports different fanouts. The fanout refers to the number of messages sent by each compute node. The fanout size is a configurable algorithmic parameter for better controlling the number of sent messages, better utilization of the network resources, and customizing the pattern based on the interconnect. 

3. We show that the Butterfly BFS communication pattern is applicable to both top-down and bottom-up BFS formulations. Specifically, the traversal operation and communication are two separate and independent phases in the algorithm. Therefore, our communication pattern can work with both types of traversals. 

4. We show a tight memory bound for Butterfly BFS and its intermediate buffers. The bound on the memory requirements means that each iteration uses the same amount of memory and that the allocation of buffers in advance is possible, resulting in fewer system calls throughout the execution.


\subsection*{Performance Contributions}

    1. We show that Butterfly-BFS scales with the introduction of additional GPUs, which is nontrivial for many multi-node BFS algorithms. In most cases, we see that we achieve a $75\%$ utilization of the peak scalability. For the 16 V100 GPUS found in the NVIDIA DGX-2 server, we can get a 12X speedup over a single GPU execution.
    
    2. Our current implementation of Butterfly-BFS uses a top-down traversal. When we compare our implementation with the direction-optimizing BFS implementation algorithm in GapBS \cite{beamer2015gap, beamer2012direction}, which is one of the fastest known BFS implementations for shared-memory systems, our new algorithm is anywhere from $2\times - 22\times$ faster, and in most cases our algorithm is over $6 \times$ faster.
    
    3. When we compare our algorithm with a top-down BFS algorithm in GapBS, our algorithm is $2\times - 233\times$ faster. For most inputs, the performance increase was in the range of $10\times - 25\times$. The speedups' difference is due to the faster performance of direction-optimizing, which avoids traversing a majority of the edges in the graph. This seems to be a promising optimization for our algorithm as well.
    
    4. For a Kronecker network with a scale of 29 (512 million vertices) and edge factor of 8 (for a total of 8 billion directed edges for the symmetrized version), our new algorithm traverses the graph at a rate over 300 GTEP/s where GTEP/s is the number of traversed edges per second in the billions. This is a high traversal rate for a single system.

    5. Our new algorithm is faster than other multi-GPU algoriths. These algorithms use both the direction optimizing traversal and use an advanced graph partitioning mechanism. In contrast, our algorithm uses a top-down traversal and a naive partitioning. Yet, our algorithm is over $50\times$ faster than the other algorithms for an identical number of GPUs even when the partitioning time is excluded from the total runtime of those algorithms.  

\section{Related Work}
\label{sec:related}

Breadth-first search, BFS for short, is a fundamental primitive used by many graph algorithms. BFS does not define the traversal order of the vertices and edges. 
BFS is applicable to both directed and undirected graphs. In undirected graphs, a BFS from a given root finds the root's connected component. BFS traversals are an integral part of many single-source shortest path (SSSP) and all-pairs shortest path (APSP) type of problems. 

BFS has received a significant amount of attention over the last two decades, and numerous optimizations are applied to make it more efficient. In the following subsections, we cover many of these optimizations. 

\paragraph*{\bf Preliminaries} 

We define a graph, $G=(V, E)$ to include the vertices $V$ in the graph and the edges $E$. An edge $(u,v)$ connects vertex $u$ to vertex $v$ in a directed manner. In an undirected graph, both $(u,v)$ and $(v,u)$ appear in the graph. The BFS traversal starts with a single vertex known as the root, typically denoted by $r$. The distance of $r$ to all remaining vertices is the output of the BFS traversal. If a vertex $u$ does not have a path to $r$ then its  distance to $d[u]=\infty$. If a vertex does have a path, then the distance $d[v]$ will represent the number of hops away.

Levels in a BFS traversal are a collection of vertices that are equidistant to the root. For parallel BFS algorithms, levels can highlight the execution parallelism. All the vertices in a given level can be traversed concurrently without worrying about finding vertices that are more than one level separated. The current traversal level has become known as the ``active-frontier'' or ``frontier''. 
Many recent algorithms use this jargon, as do we.

\paragraph*{\bf Single compute-node BFS}

Alg. \ref{alg:pseudo-bfs} depicts pseudo-code for a parallel BFS typically found in shared-memory systems. Distributed BFS algorithms are different and require a global communication phase and synchronization.
Thus, Alg. \ref{alg:pseudo-bfs} is a good fit for shared-memory systems (or single compute-node systems) as its implementation is relatively easy with either OpenMP (CPU) or CUDA (GPU). 

While the pseudo-code for of Alg. \ref{alg:pseudo-bfs} is simple, achieving high performance is non-trivial due to numerous challenges, including load-balancing, system utilization, memory access patterns. The main observation from Alg. \ref{alg:pseudo-bfs} is that all vertices at a specific level and all their respective edges get traversed concurrently.
The next level of the BFS gets discovered in each iteration of the $While-loop$. The distance from the root to all new vertices is also updated. When extending the algorithm from one compute-node to multiple compute-nodes, these two arrays are synchronized across the network to ensure that all compute-nodes have a correct view of the newly found vertices.

\begin{algorithm}[t]

\tiny

\caption{Parallel Top Down BFS. $Q$ containts two queues - one for the current frontier and one for the next frontier. }
\label{alg:pseudo-bfs}

\ForPar{$v=0:1:|V|$}
{
$  d[v]\leftarrow infty$ \\
}

$level \leftarrow 0$ \\
$Q.Enqueue(root)$ \\ 
$d[root] \leftarrow 0$ \\

\While{$Q.NotEmpty()$}
{
	\ForPar{$v\in Frontier$}
	{
		\ForPar{$u \in adj(v)$}
		{
			\If{$d[u]=\infty$}
			{
				$d[u] \leftarrow d[v]+1 $\\
				$Q.Enqueue(u)$ \\
			}
		}
	}
	$Q.SwapQueues()$  // Replace current frontier with next frontier \\
	$level \leftarrow level + 1$ \\
}

\end{algorithm}

\paragraph*{\bf Top Down Vs. Bottom Up}

Beamer \emph{et al.} \cite{beamer2012direction} introduce direction optimizing BFS, where at run-time, the algorithm can choose one of two traversal mechanisms. Direction optimizing BFS uses both the classic \emph{top-down} traversal and a new \emph{bottom-up} traversal. The classic top-down traversal is where a parent vertex finds all its children in the level below it. In the new bottom-up traversal, children find their parents and verify that they are part of the current frontier. The child-parent lookup gets conducted across all the vertices in the graph. This optimization is especially effective for ``small-world networks'' where the number of levels is relatively small and most vertices get discovered in two or three iterations. For such networks, it is possible to avoid traversing a majority of the edges.

In contrast, networks with large diameters do not benefit as much from the direction-optimizing traversal as either the bottom-up traversal can add overhead or a majority of the time is spent doing the top-down traversal.  
The increase in popularity of the Graph500 benchmark had led many high-performance BFS algorithms to adopt the direction-optimizing traversal. Specifically, the Graph500 benchmark uses a graph generator that creates small-world networks. The Graph500 reports performance in TEP/s - Traversed Edges Per second. 
As direction optimizing BFS is great for small-world networks, its adoption is not a big surprise. In practice, the number of traversed edges per second gets reported as the number of edges in the graph divided by the execution time. Thus, the above TEP/s metric is not accurate and does not represent the number of traversed edges or random memory accesses as  $90\%$ of the edges can be avoided due to the direction optimizing traversal.

Lastly, direction optimizing BFS does not apply to all problems requiring a BFS traversal. For example, an APSP type of problem such as betweenness centrality might need to find all paths. Other examples include weight-filtering BFS where only edges with a given weight are traversed.

\paragraph*{\bf Parallel \& Distributed BFS}

Barrett \emph{et al.} \cite{barrett2009implementing} and Madduri \& Bader \cite{bader2006designing} show that parallel BFS can be efficient on shared-memory systems that have a good amount of fine-grain parallelism. The algorithm by Madduri and Bader \cite{bader2006designing}  is similar to the algorithm presented in Alg. \ref{alg:pseudo-bfs}. Many shared-memory algorithms assume that the system memory has a fairly uniform memory access time. This is in stark difference to distributed algorithms.

Dhulipala \emph{et al.} \cite{dhulipala2018theoretically} show that shared-memory systems can process extremely large graphs (with some 200 billion edges). Their solution uses three different frameworks: LIGRA \cite{shun2013ligra}, LIGRA+ \cite{shun2015smaller}, and Julienne \cite{dhulipala2017julienne}. LIGRA+ adds a graph compression layer over LIGRA. The largest graph used by Dhulipala \emph{et al.} \cite{dhulipala2018theoretically}  is 900GB, and it gets compressed to 330GB. The size reduction also reduces the amount of data fetched into memory and the caches. 

Yoo \emph{et al.} \cite{yoo2005scalable} show a scalable BFS algorithm design for the BlueGene/L system. Specifically, customized communication collectives were designed to utilize the torus interconnect. The paper also covers additional optimizations such as 2D partitioning, which can help reduce the number of messages from $P$ to $\sqrt{P}$ where $P$ is the number of processors. This algorithm assumes that the adjacency matrix is dense, which is not common in many real-world graphs.

Buluc and Madduri \cite{bulucc2011parallel} show a scalable distributed BFS that utilizes a sparse graph data structure and uses a BLAS (Basic Linear Algebra Subprogram) based formulation. The BLAS operations in \cite{bulucc2011parallel} are on sparse vectors and matrices. Beamer \emph{et al.} \cite{beamer2013distributed} extend the algorithm by Buluc and Madduri \cite{bulucc2011parallel} to use the direction optimizing formulation. The direction optimizing formulation improves performance by roughly $7\times$.

Lin \emph{et al.} \cite{lin2018shentu} show a new scalable graph framework, ShenTu, designed for massive graphs for the Taihu system. ShenTu scales to graphs with trillions of vertices and edges. The Emu Chick \cite{dysart2016highly} is a fairly new architecture designed for sparse applications. 
Several recent studies have studied graph algorithms and BFS (in particular) on the Emu Chick. Hein \emph{et al.} \cite{hein2018initial} review several different approaches to implement random memory accesses. Hein \emph{et al.} \cite{hein2018programming} and Belvirani \emph{et al.} \cite{belviranli2018designing} study different BFS algorithms on the Emu Chick. Hein \emph{et al.} \cite{hein2018programming} show that a BFS algorithm using remote writes is more efficient than using thread migrations. Belvirani \emph{et al.} \cite{belviranli2018designing} show that an Intel processor outperforms the Emu Chick while the Chick shows good scalability.

\paragraph*{\bf GPU Accelerated BFS}
Merrill \emph{et al.} \cite{MerrillGPU} showed the first GPU accelerated BFS implementation on the GPU. 
The Gunrock framework \cite{wang2015gunrock} has several highly tuned BFS implementations, including a direction optimizing BFS implementation that swaps between the classic top-down and bottom-up algorithms. Gunrock allows users to configure many run-time parameters, including the load-balancing scheme and queuing mechanism. 
The Hornet framework \cite{green-hornet} also has a direction optimizing BFS traversal implementation, though it does not offer these run-time parameters. 

Busato and Bombieri \cite{busato2018efficient} evaluate several load-balancing mechanisms for graph traversals on the GPU. These load-balancing mechanisms work at various parallel granularities, ranging from the warps and all the up-to-the whole device. 
Logarithmic Radix Binning  (LRB) is a load-balancing mechanism first introduced in \cite{green-tri-lrb-gpu,green-tri-lrb} for triangle counting and later generalized for other graph problems and segmented sorting \cite{green-lrb-2019}. Our ButterFly BFS algorithm uses LRB for performance purposes. 
We cover Hornet in additional detail in Section \ref{sec:hornet} as our new algorithm works within the Hornet framework.

\paragraph*{\bf Multi-GPU BFS}
\label{sec:multi-gpu}
The first multi-GPU systems were designed such that the numerous GPUs were placed within the same server. Communication between the GPUs occurred over the PCI-E interconnect with two typical communication types: peer-to-peer communication or through global managed memory shared by the GPU. The key difference between these two communication methods is that in the peer-to-peer communication approach, the CPU manages the peer-to-peer messages, and each GPU is responsible for managing its memory and will only access its memory. 
In contrast, the managed memory method uses a GPU internal software framework that moves data between multiple devices by triggering page-faults. The first approach is closer to a traditional message-passing system as each GPU is responsible for its memory and communicates with the other GPUs by messages. The latter approach is closer to a shared-memory approach such that all the devices can view the memory of all devices. While the second approach can be convenient, it can also come at the cost of performance due to massive data migration.
Furthermore, having multiple GPUs share the same PCI-E leads to contention on the interconnect. 

To put things into perspective, the NVIDIA A100 and V100 GPUs have a bandwidth of 1,555 $GB/s$ and 900 $GB/s$, respectively. In contrast, PCI-E V4 and PCI-E V3 with 16 channels have a bandwidth of 32$GB/s$ and 16 $GB/s$, respectively. Thus, the cost of communication across the slow PCI-E interconnect can significantly reduce performance.
In most cases, the network bandwidth is also one to two orders smaller than the on-device memory. For this reason, avoiding unnecessary data movement and controlling communication can be very beneficial to the performance of a multi-GPU algorithm. 
The ratio between on-device and peripheral bandwidth is similar to the ratio between on-device bandwidth and network bandwidth (such as Infiniband or Ethernet). For this reason, some of the original multi-GPU servers resemble a small cluster of servers with a single GPU inside of them. The introduction of NVIDIA's NVLink interconnect slightly changes this view.

NVIDIA's NVLink interconnect responsible for connecting multiple GPUs has made a big difference in allowing more data movements with lower overheads. Many multi-GPU systems have dedicated channels between the GPUs, allowing for fast and direct contact. The NVIDIA A100 and V100 GPU support up to 12 NVLink channels. 
In the NVIDIA DGX A100, a single A100 GPU can have up to 600 GB/s of bandwidth to the remaining GPUs. 
In the NVIDIA DGX-2, a single V100 GPU  can have up to 300 GB/s of bandwidth to the remaining GPUs. 
NVIDIA's NVSwitch technology available in the NVIDIA DGX A100 and DGX-2 allow for direct GPU to GPU communication between all the devices.
These peak bandwidths are not meant for small random memory accesses but instead designed for mid-size to large messages. Nonetheless, the introduction of these technologies opens up new opportunities for algorithm designers. 
 
Our algorithm is designed to utilize NVLink and NVSwitch within a single DGX server using both the high-bandwidth and the point-to-point communication. At the same time, our algorithm goes with the more traditional message-passing approach where each GPU is responsible for managing its communication with the other devices. While global shared-memory makes some implementations details easier, the side effect of non-uniform memory access (NUMA) would significantly reduce performance.

Hong \emph{et al.} \cite{hong2011efficient} show a GPU and CPU hybrid algorithm that decides if the execution of a given frontier traversal should be on the GPU or CPU. To avoid transferring the graph when swapping execution,  the CPU and GPU each require a copy of the input. 
A similar approach is taken by Gharaibeh \cite{gharaibeh2012yoke}. The approach of sharing the computation between the CPU and GPU is also found in recent studies, such as accelerating betweenness centrality using multiple GPUs \cite{mishra2020fine}. Mishra \emph{et al.} conducted their experiments on older generation CPUs and GPUs. Communication between the CPU and GPU was through a slower PCI-E interconnect. The bandwidth gap between the on-device memory and the interconnect between the CPU and GPU makes the hybrid approach undesirable. 

Groute \cite{ben2017groute} is an asynchronous framework designed to make scheduling and routing decisions for multi-GPU graph analytics. It does so by sending small messages to the devices responsible for executing the traversal. Groute outperforms the multi-GPU graph algorithms in Gunrock \cite{pan2015multi}. Yet, both Groute and Gunrock show limited scalability going from one GPU to multiple GPUs.  
A more recent study by Jatala \emph{et al.} \cite{jatala2020study} shows that the performance of several distributed frameworks for BFS, including the asynchronous Gluon \cite{dathathri2019gluon}, across multiple configurations. The key execution bottleneck for these algorithms is communication, which accounts for roughly $70\%$ of the total execution time. In contrast, our algorithm spends a small portion of its time in the communication phase due to the butterfly communication pattern and the fast interconnect.

\section{ButterFly BFS}
\label{sec:butterfly}

In this section, we introduce our new ButterFly BFS algorithm. Specifically, we show how to use a butterfly communication network to synchronize the distance arrays and queues across the GPUs. The butterfly network replaces the typical all-to-all communication pattern. The butterfly communication pattern reduces the number of messages and network traffic at the cost of doing several rounds of communication. Furthermore, the butterfly communication pattern allows placing an upper bound on the size of the messages and memory buffers whereas the
all-to-all communication pattern does not offer these bounds.

\paragraph*{\bf Single compute-node to Multiple compute-node Challenges }

There are numerous challenges associated with scaling a BFS traversal from one compute-node to multiple compute-nodes. These problems include load-balancing, partitioning, data synchronization, and communication. Give that anyone of the compute-nodes can find any vertex $v$ in the graph as part of the traversal requires the compute-nodes to ``communicate and synchronize'' the set of vertices found in this current level of the BFS. Without this synchronization, compute-nodes cannot know which vertices (regardless of vertex ownership) are part of the current or next frontier. Thus, after each compute-node has completed the edge traversal process, a global synchronization across all compute-nodes is necessary.

Given $P$ compute-nodes, the data on each compute-node is disseminated to the remaining $P-1$ compute-nodes\footnote{We distinguish between threads and compute-nodes by the number of communicators. For example, a multi-threaded CPU system might be a single compute-node for an OpenMP-based implementation, or it might be considered multiple compute-nodes for an MPI-based implementation where each thread has a communicator, or its own rank. A single GPU has thousands of threads, yet it is a single compute-node as it requires a single communicator.}. Thus, all-to-all communication is necessary for synchronization.
At the heart of this challenge is the question of how to implement the all-to-all communication efficiently. Two widely used and naive approaches for this all-to-all include
 (additional discussion in Sec. \ref{sec:related})
: 1) each compute-node sends its data to the remaining GPUS in a concurrent manner, and 2) each compute-node communicates with the remainder of the compute-nodes, one at a time in an iterative fashion. 
Both approaches send an equal amount of data and the same number of messages, roughly $O(P^2)$ messages.
The first of these approaches supposedly has lower latency as it does its communication in a bulk manner. However, it also typically saturates the communication network between the nodes and can cause congestion due to a large amount of data and number of messages. The first approach also requires having a large buffer available to store the incoming messages; this has an upper-bound of $O(E)$ elements and assumes that each GPU finds all the vertices in the graph at once. In contrast, the second approach has a reduced footprint of $O(V)$, which is the maximal number of vertices. The second approach requires $P-1$ rounds of communication. 
This problem becomes more prevalent as $P$ scales to a more significant number of compute-nodes.

\begin{figure*}[t]
\centering

\subfloat[Butterfly network with fanout of one.]{\includegraphics[width=0.31\columnwidth]{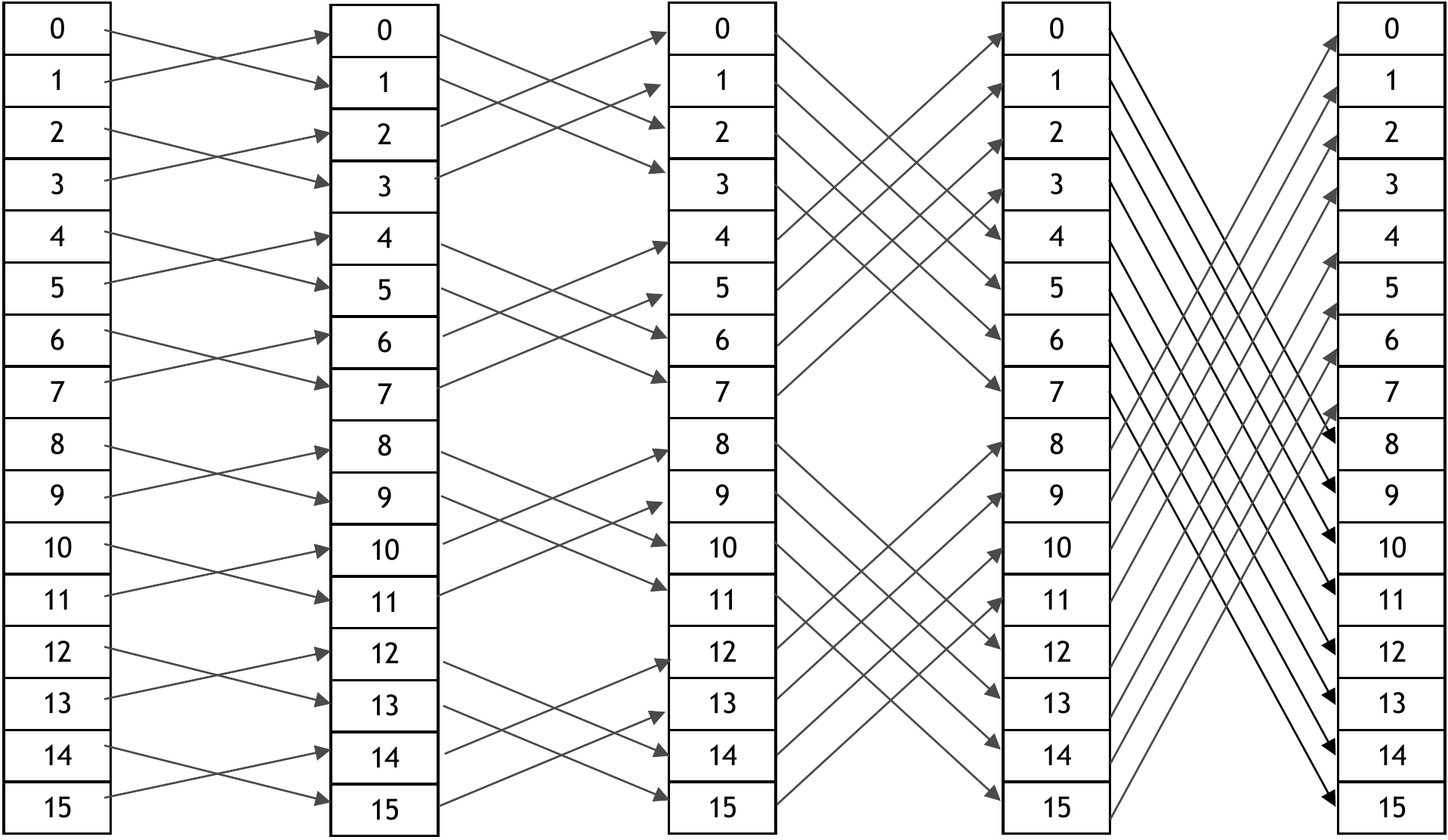}}
\hspace{0.4cm}
\subfloat[Queue after local traversal.]{\includegraphics[width=0.31\columnwidth]{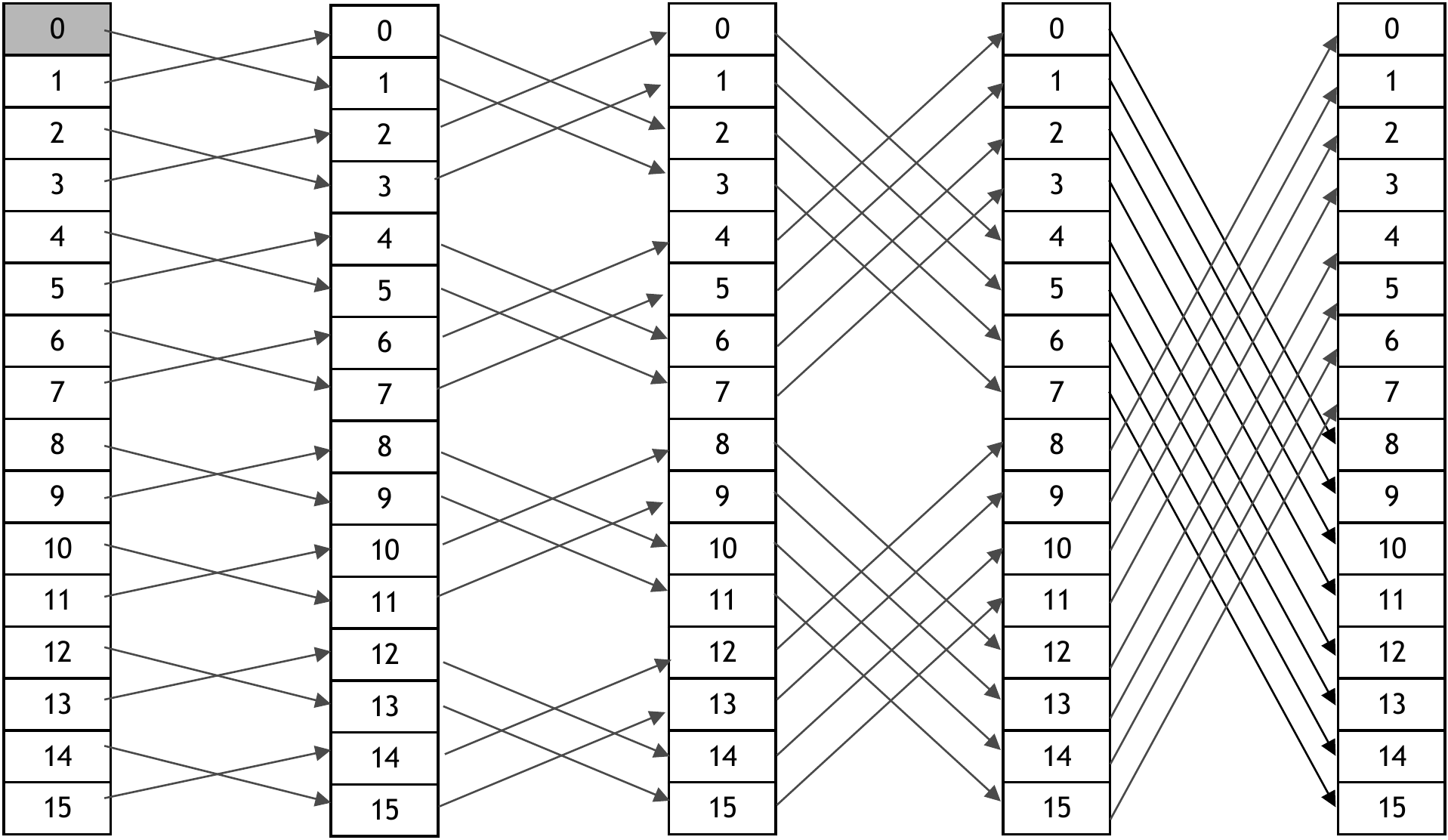}}
\hspace{0.4cm}
\subfloat[Queue after one round of communication.]{\includegraphics[width=0.31\columnwidth]{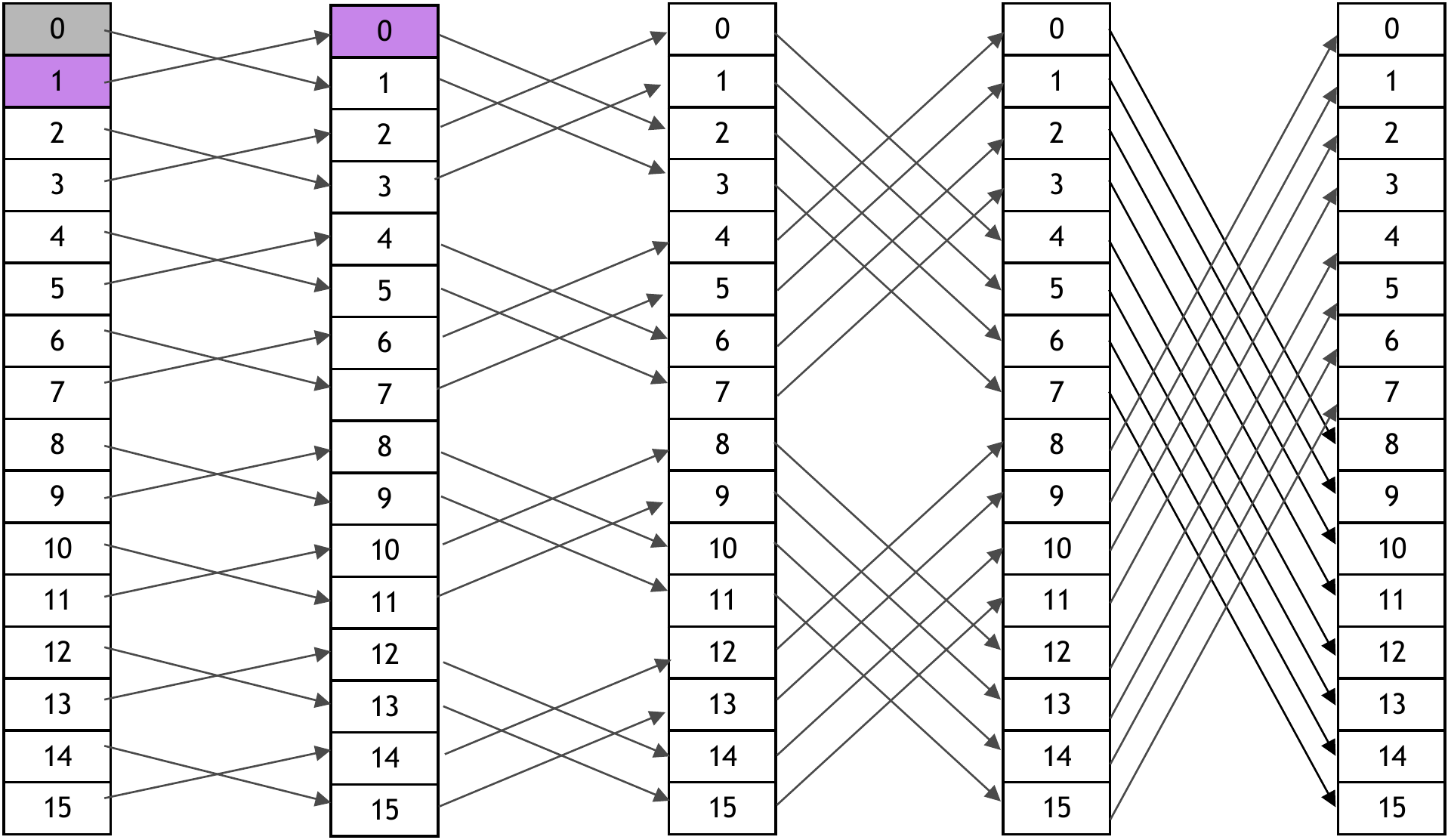}}

\subfloat[Queue after two rounds of communication.]{\includegraphics[width=0.31\columnwidth]{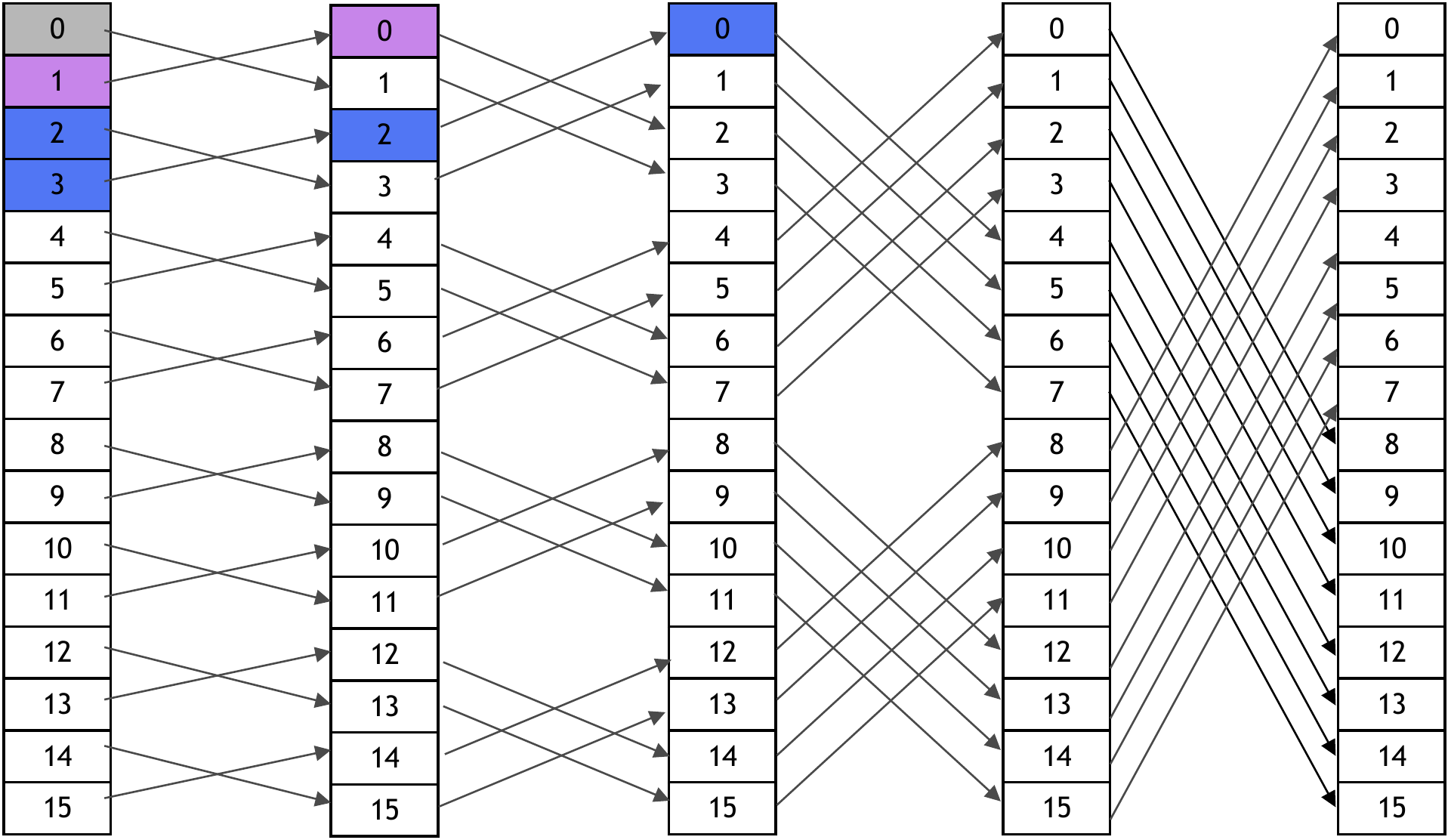}}
\hspace{0.4cm}
\subfloat[Queue after three rounds of communication.]{\includegraphics[width=0.31\columnwidth]{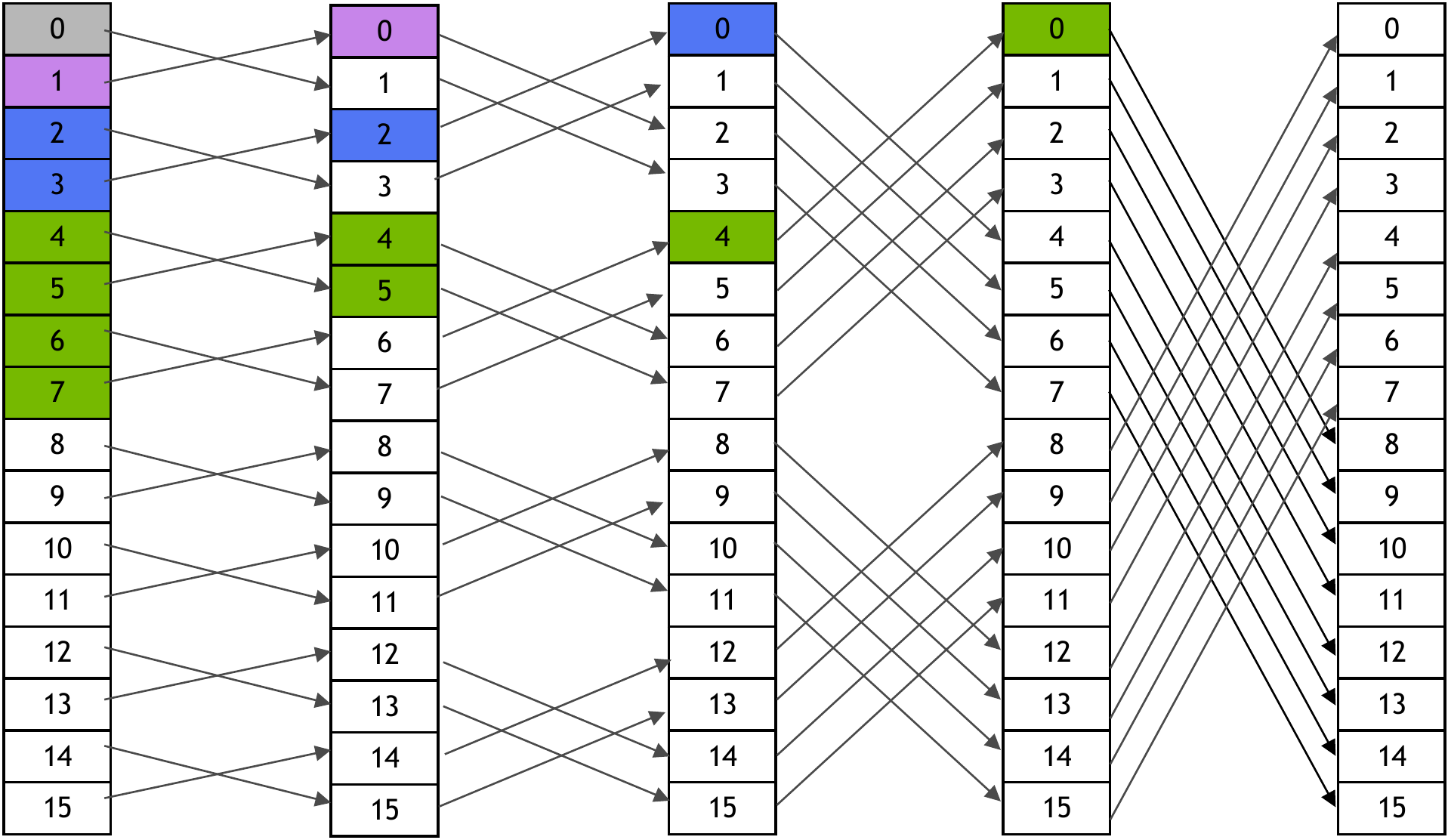}}
\hspace{0.4cm}
\subfloat[Queue after four rounds of communication.]{\includegraphics[width=0.31\columnwidth]{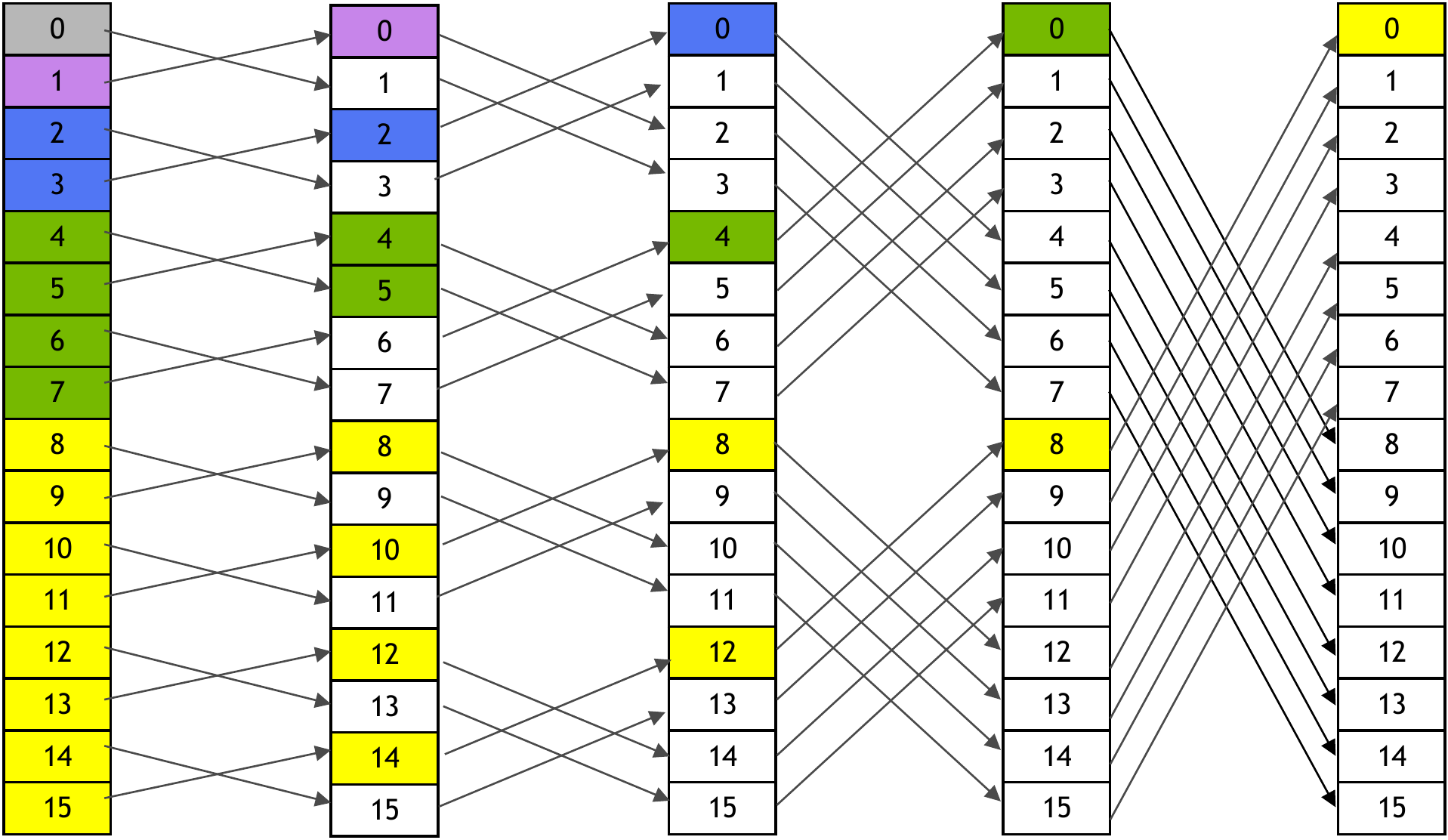}}

\caption{Depicts the butterfly network with a fanout of one for sixteen GPUs as a function of the communication rode. We focus on compute node 0.}

\label{fig:fanout1}
\end{figure*}

\begin{figure*}[t]
\centering

\subfloat[Butterfly network with fanout of four.]{\includegraphics[width=0.220\columnwidth]{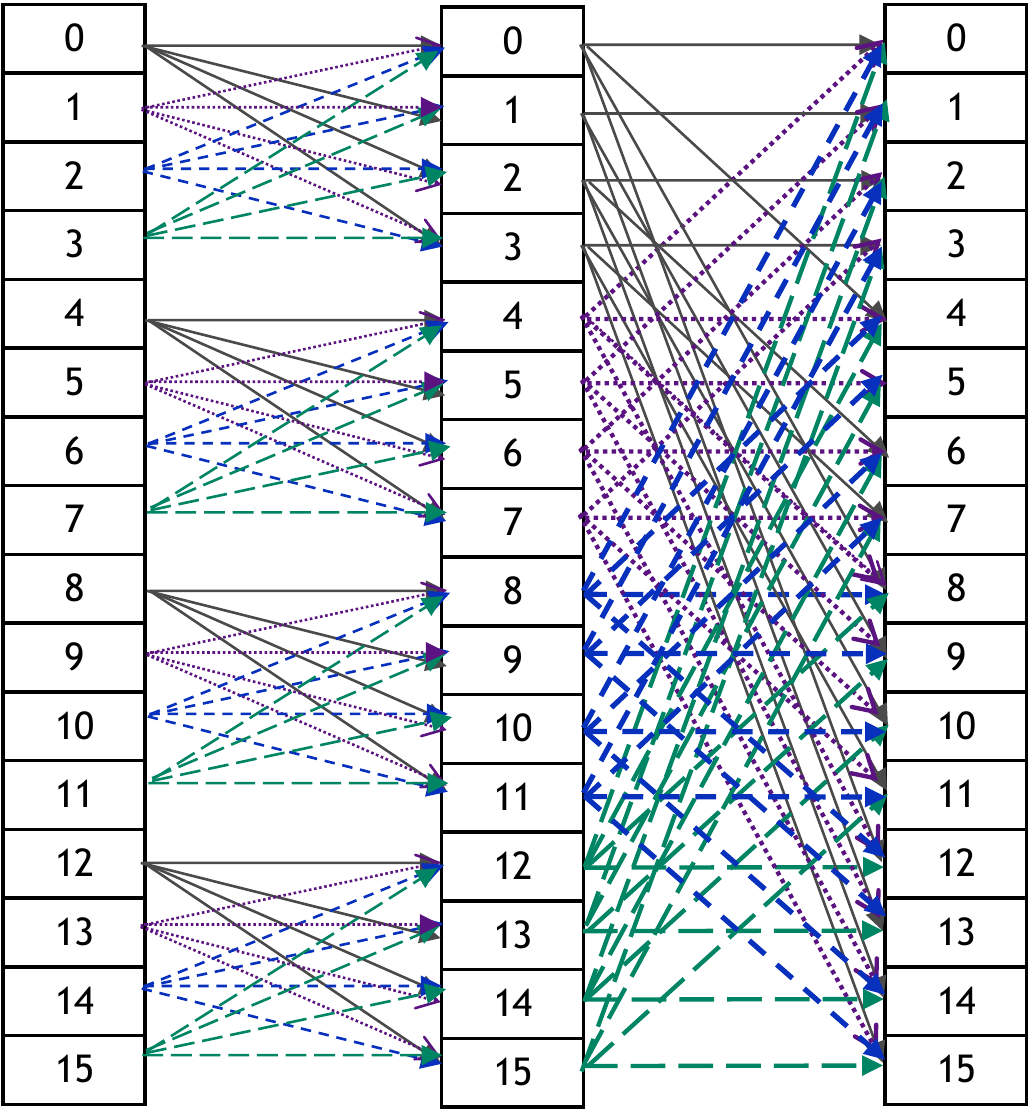}}
\hspace{0.4cm}
\subfloat[Queue after local traversal]{\includegraphics[width=0.220\columnwidth]{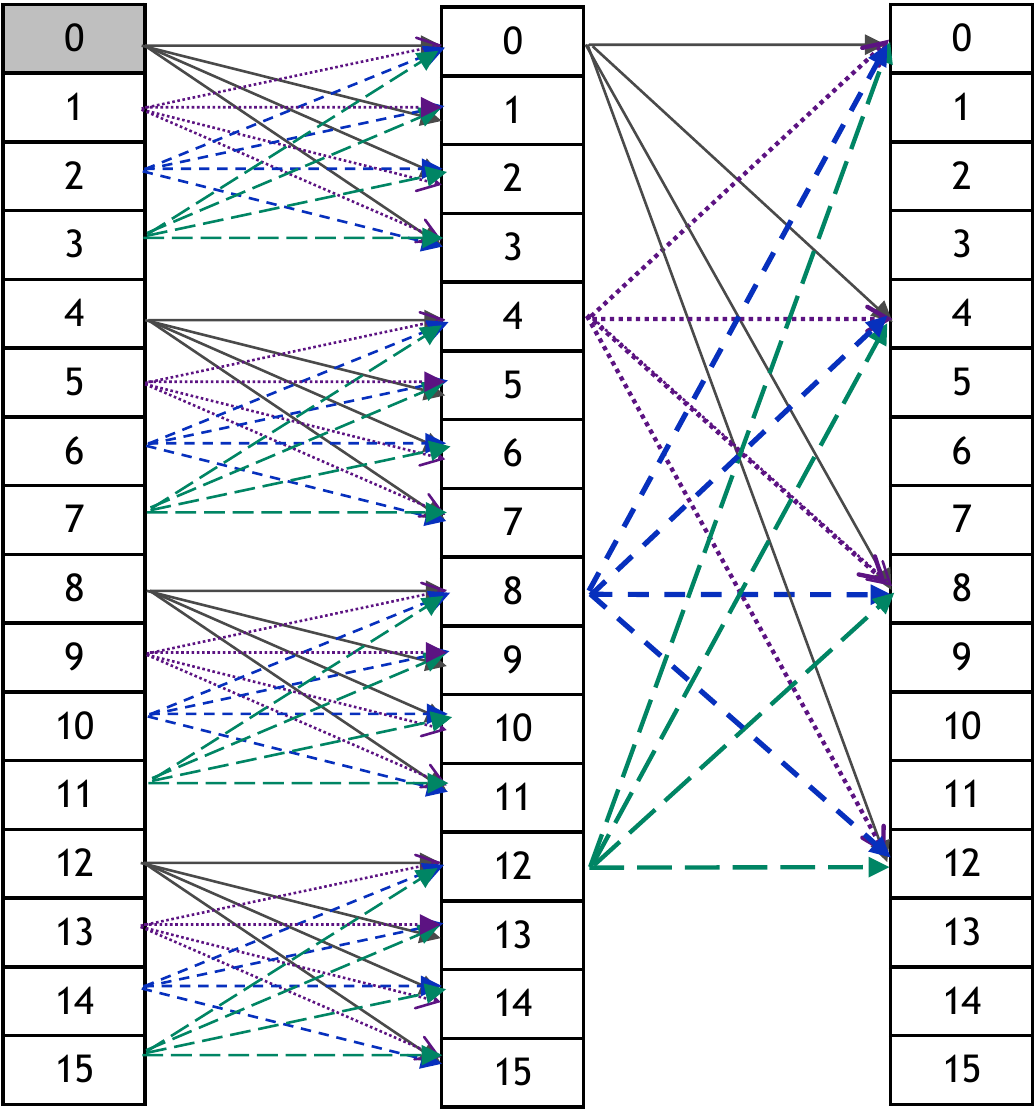}}
\hspace{0.4cm}
\subfloat[Queue after one round of communication.]{\includegraphics[width=0.220\columnwidth]{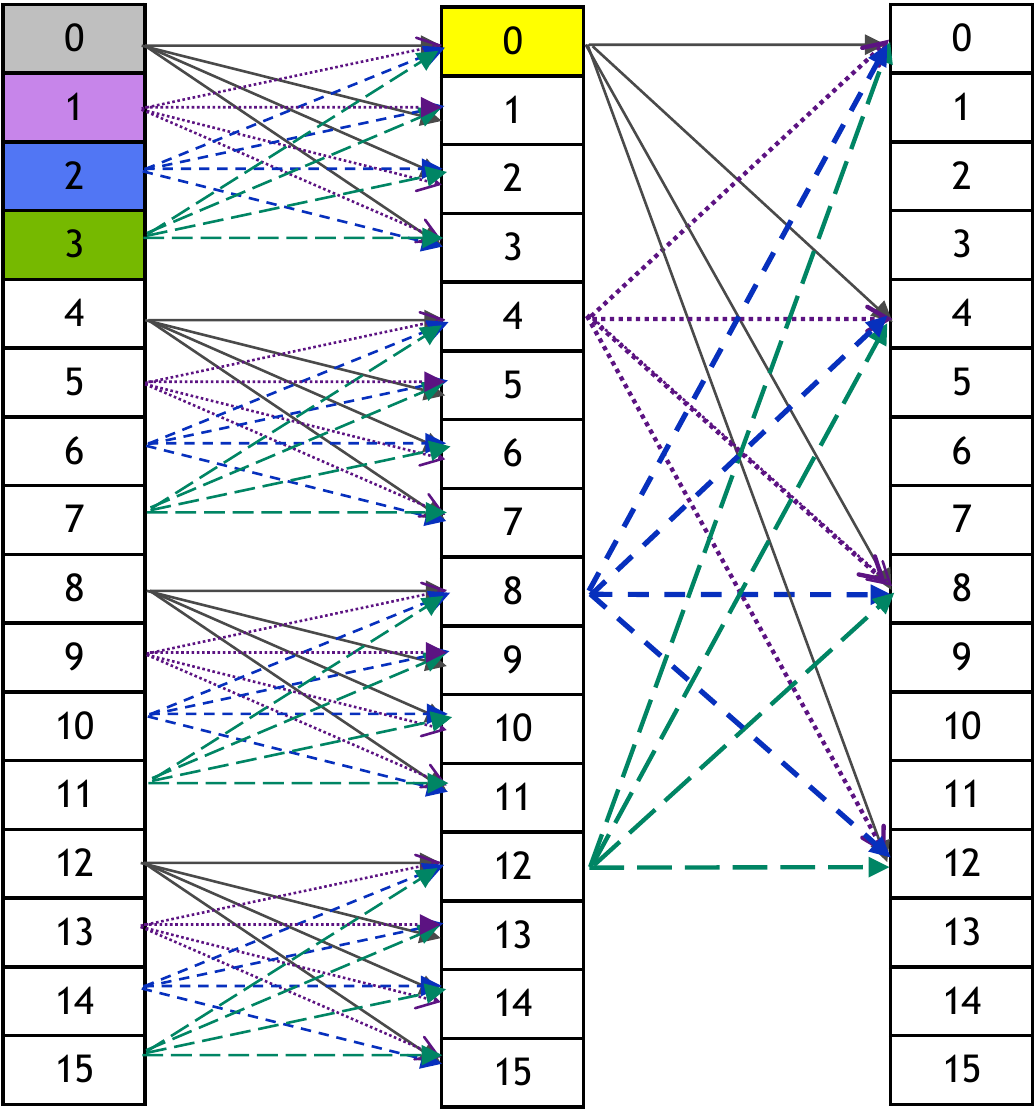}}
\hspace{0.4cm}
\subfloat[Queue after two rounds of communication.]{\includegraphics[width=0.220\columnwidth]{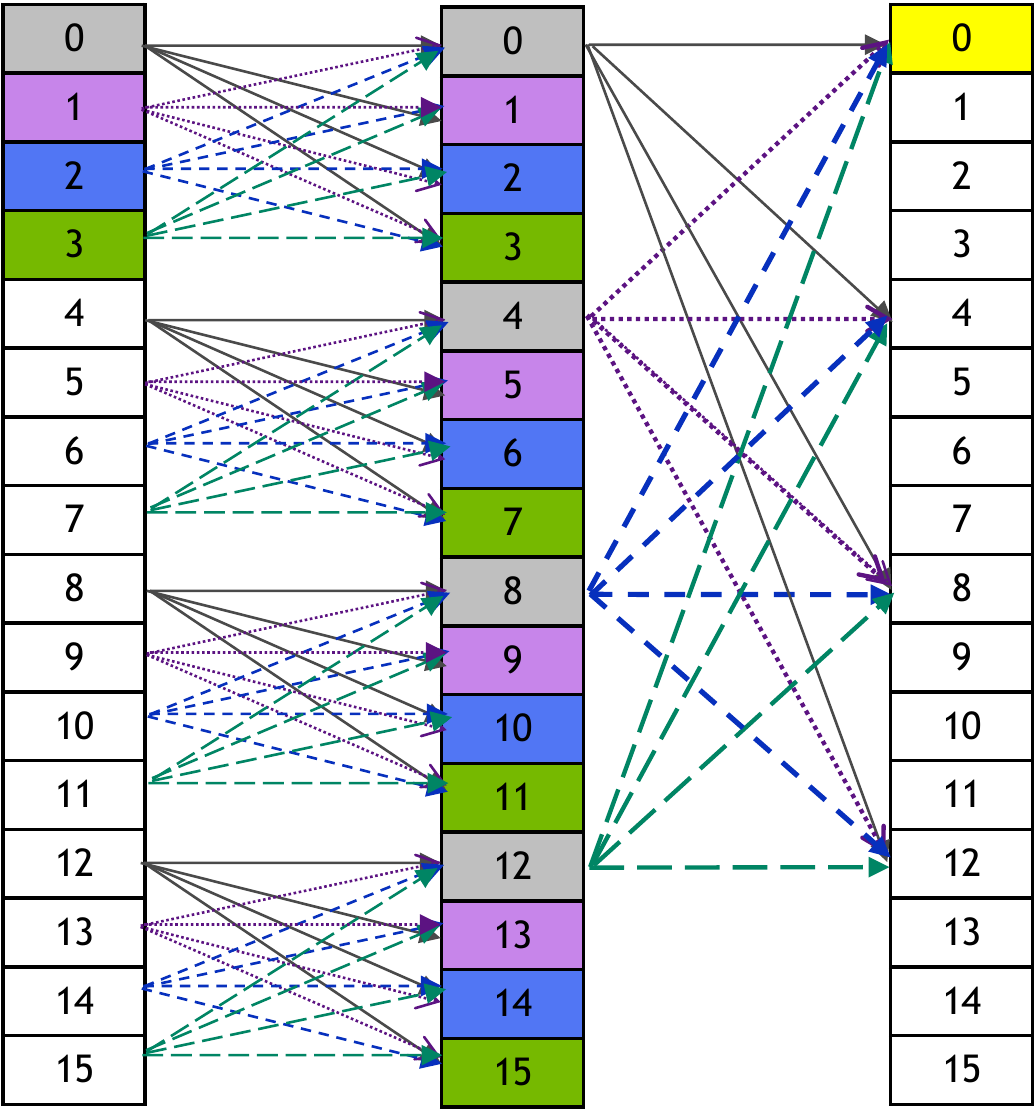}}

\caption{Depicts the butterfly network with a fanout of four for sixteen GPUs as a function of the communication rode. }

\label{fig:fanout4}
\end{figure*}

\begin{algorithm}[t]
\DontPrintSemicolon

\tiny

\caption{ButterFly BFS pseudo-code. $CN$ refers to Compute Nodes. For simplicity, this pseudo-code is for fanout of one. }

\label{alg:pseudo-butterfly-bfs}

\ForPar{$g=0:1:|CN|$}{
	\ForPar{$v=0:1:|V|$}
	{
	$  d_{local}[g][v]\leftarrow infty$ \\
	}	
}

$level \leftarrow 0$ \\

\ForPar{$g=0:1:|CN|$}{
	\If{$root \in myVertics[g]$}
	{
		$Q_{local}[g].Enqueue(root)$ // Only one CN will enqueue the root //
	}
	$d_{local}[g][root] \leftarrow 0$ // All CN set their d
}

\While{$Q.NotEmpty()$}
{
	{\bf Phase 1: Traversal}\\

	\ForPar{$g=0:1:|CN|$}{

		\ForPar{$v\in Q_{local}[g$}
		{
			\ForPar{$u \in adj(v)$}
			{
				\If{$d{local}[g][u]=\infty$}
				{
					$success \leftarrow Q_{global}[g].Enqueue(u)$ // Requires an atomic operation for multiple threads per rank\\
					\If{$success$} {
						$d_{local}[g][u] \leftarrow level+1 $\\
						\If{$u \in myVertices[g]$}{
							$Q_{local}.Enqueue(u)$ \\
						}
					}
				}
			}
		}
	}

	{\bf Phase 2: Butterfly Network \& Frontier Synchronization}\\

	\ForPar{$g=0:1:|CN|$}{
		\For{$i=0:1:log_{2}(|CN|)$}{		
			$srcCN \leftarrow ButterflyDirection(g,i)$ // Get CN that will be transferring frontier \\
			$tempFrontier \leftarrow CopyFrontier(Q_{global}[srcCN])$

			\ForPar{$v \in tempFrontier $}{		
				\If{$d_{local}[g][v] = \infty$}{

					$Q_{global}[g].Enqueue(v)$ \\
					$d_{local}[g][v] \leftarrow level+1 $\\

					\If{$v \in myVertices[g]$}{
						$Q_{local}[g].Enqueue(v)$ \\
					}
				}
			}
			$Synchronize()$
		}
	}


	$Q.SwapQueues()$  // Replace current frontier with next frontier \\
	$level \leftarrow level + 1$ \\
}


\end{algorithm}

\paragraph*{\bf Synchronizing with ButterFlies}

Given the communication challenges above, we introduce a new communication pattern for BFS that reduces the number of messages sent over the communication network. Specifically, we show that a {\bf butterfly network} can be used to reduce the number of messages and the amount of data sent. We refer to this network as a butterfly network as it resembles the butterfly network used by Cooley and Tukey \cite{cooley1965algorithm} in their seminal paper on fast FFTs. We show that a similar communication pattern allows synchronizing the frontiers across the compute-nodes. This is the first time this communication pattern has been applied to distributed breadth-first search to the best of our knowledge.

In the following, we cover in additional detail our novel algorithm. We use the pseudo-code in Alg. \ref{alg:pseudo-butterfly-bfs} with  Fig. \ref{fig:fanout1} and Fig. \ref{fig:fanout4} to explain how our Butterfly BFS algorithm works. 
As a starting point, note that Alg. \ref{alg:pseudo-butterfly-bfs} assumes that there are multiple compute-nodes (denoted as $CN$ in the pseudo-code), this is in contrast to Alg. \ref{alg:pseudo-bfs} which assumes that only one compute-node exists. Adding more compute-nodes to the computation now requires that the compute-nodes synchronize the vertices of a new frontier to ensure that the next frontier is found correctly on each of the compute-nodes.

Fig. \ref{fig:fanout1} (a) shows how to synchronize the frontiers using a butterfly-network. Fig. \ref{fig:fanout4} extends the butterfly network to use a fanout of four (covered in additional detail in \ref{sec:sub_fanout}). The butterfly communication patterns shows how the compute-nodes synchronize the frontiers. For simplicity, we show how the data from all the compute-nodes gets transferred to compute-node 0---this process is explained through the subfigures labeled (b)-(f). A similar process occurs for the remaining compute-nodes.
Fig. \ref{fig:fanout1} (b) depicts the fact that compute-node 0 is only aware of the vertices in it has found. Fig. \ref{fig:fanout1} (c) illustrates that compute-node 0 has synchronized its frontier with that from compute-node 1. 
Fig. \ref{fig:fanout1} (d) depicts that compute-node 0 has synchronized its frontier with that from compute-node 2 and compute-node 2 has the frontiers from compute-nodes 2 and 3. At this point, compute-node 0 has the frontiers of four different compute-nodes. In the next phase, Fig. \ref{fig:fanout1} (e), compute-node will receive the frontier of compute-node 4, which has the frontier of compute-nodes 4-7. Lastly, Fig. \ref{fig:fanout1} (f) depicts the synchronization with compute-node 8 and its frontier of compute-nodes 8-15. Thus, in four steps, compute-node 0 has received information from 16 compute-nodes. Note, the goal of Fig. \ref{fig:fanout1} is to show how the data is propagated between the compute-nodes. The code needed for the frontier synchronization is captured in Alg. \ref{alg:pseudo-bfs}.

\paragraph*{\bf Frontier Propagation}

Alg. \ref{alg:pseudo-butterfly-bfs} depicts the pseudo-code for our new Butterfly BFS algorithm. Note, there are multiple tiers of parallelization within the algorithm. The first tier of parallelization is at the distributed system level, where the work gets split across multiple compute-nodes. 
Here, the compute-nodes work in a synchronized manner and that there are implicit synchronizations at the end of the parallel $for-loops$ over the compute-nodes. 
The second tier of parallelization is at the compute-node granularity. 
Specifically, additional parallelism within the compute-node itself is now possible. 

Another observation regarding Alg. \ref{alg:pseudo-butterfly-bfs} is that the pseudo-code makes several simplifying assumptions: a) the number of compute-nodes is a power of 2 and that the b) the fanout is one. Overcoming these two assumptions is fairly easy but adds some complexity to the pseudo-code itself. We chose simplicity for presentation purposes but note that the algorithm is not constrained to the above configuration. Furthermore, to simplify computing the edges of the butterfly communication network, we use a function denoted as {\it ButterflyDirection()} that computes the source of the communication based on the destination compute-node and the iteration within the butterfly network. 

Note that Alg. \ref{alg:pseudo-butterfly-bfs} has two very distinct phases. The first phase is responsible for traversing the edges of the current frontier on each of the compute-nodes. This phase is similar to the single compute-node algorithm found in Alg. \ref{alg:pseudo-bfs}, with the big distinction that the parallelization is now across multiple compute-nodes. Furthermore, this multi compute-node algorithm requires two queues for each compute-node: a global queue and a local queue. The global queue on each compute-node stores all the vertices found on that compute-node as part of the current frontier traversal. It is the global queue that is used for synchronizing the frontiers in the butterfly communication network. The second queue stores all the vertices found in the current frontier traversal that belong to a specific compute-node. Recall, each vertex belongs to precisely one compute-node and will only be in one local queue.  In the next BFS frontier, these are the active vertices of a given compute-node. 

The second phase, the butterfly communication network, is where the vertex frontiers are propagated to the remaining compute-nodes, and results in updating of the local queues. The receiving compute-nodes get lists of active vertices of a frontier, they then proceed to check if a given vertex already exists within their copy of the global queue. If the vertex doesn't exist, then it is added to the global queue. The global queue might be part of an additional round of the butterfly network. Lastly, the compute-node checks if the vertex belongs to its set of vertices and whether it need be added to the local queue. 
Once the butterfly communication has concluded, every compute-node has a list of the next frontier's active vertices.

\paragraph*{\bf Fanout \& Trade-offs}
\label{sec:sub_fanout}
Fig. \ref{fig:fanout1} depicts the butterfly network when each compute-node synchronizes its frontier with one compute-node in each iteration of the butterfly network---this is commonly referred to as a fanout of one. The fanout can be extended to support more communication such as a fanout of size four as depicted in Fig. \ref{fig:fanout4} where each compute-node's frontier is synchronized against the frontiers of four compute-nodes. After one iteration of communication, Fig. \ref{fig:fanout4} (c), compute-node has synchronized against compute-nodes 0-3. In the next iteration, Fig. \ref{fig:fanout4} (d), compute-node synchronize against compute-nodes 0,4,8, and 12 which have the frontiers of compute-nodes 0-3, 4-7, 8-11, and 12-15, respectively.
The structure of the fanout-4 network is similar to the fanout-1 network. There are several distinctions worth noting:
1) The depth of the communication network is smaller as the number of synchronization is reduced.
2) The number of messages sent is higher.
3) The buffer needed to store the global queues is larger. 
4) Fewer communication bottlenecks. In Fig.\ref{fig:fanout1} (f), note that compute node $8$ might need to communicate with eight different compute nodes in the last iteration if there are nine total compute nodes. This performance side-effect is visible in Fig. \ref{fig:time-bfs} for fanout one when going from eight compute nodes to nine.

In the following, we analyze the complexity of these networks. We denote the fanout as $f$. {\bf For the sake of brevity and simplicity, our analysis focuses on systems where the number of compute-nodes is a power of two. The complexity does not change by much for non-powers of two but requires a slightly more detailed analysis}.

Recall that the frontier in a BFS traversal can never be larger than $O(V)$ elements. To store the incoming frontiers of $f$ compute-nodes, a buffer of size $O(f \cdot V)$ is required (assuming a single array is pre-allocated to avoid run-time overheads). Thus, the difference in the buffer size between a fanout of $f=1$ and $f=4$ is $4\times$.

The depth of the network for a fanout of one and four is $log_{2}(CN)$ and $log_{4}(CN)$, respectively. This can be further generalized to $log_{f}(CN)$
In each iteration of the communication, each compute-node sends $f$ messages. Thus, each compute-node sends roughly $f\cdot log_{f}(CN)$ messages, and all the compute-nodes send together a total of $CN \cdot f\cdot log_{f}(CN)$ messages. For a fanout of 1 and 16 compute-nodes, a total number of 64 messages are necessary. In contrast, for a fanout of 4 and 16 compute-nodes, a total of 128 messages are needed. 
Lastly, it is possible to set the fanout $f=CN$; this is an equivalent to an all-to-all communication, which requires a $CN^{2}$ messages and potentially a larger buffer for the intermediate results.

Thus, the fanout of the communication network is a useful parameter for trading-off the storage, amount of data transferred across the data, number of messages, and the number of iterations. In practice, the decision on $f$ should take into account the number of unique hardware network controllers. Specifically, a fanout of 1 might not use multiple network cards or other types of interconnects and could result in a slower/deeper network.

\paragraph*{\bf Parallelization Schemes and Direction Optimizing}

Note, the code in Alg. \ref{alg:pseudo-butterfly-bfs} shows that the BFS traversal is also parallelizable at the compute-node granularity. For the sake of simplicity, we show pseudo-code for the classic top-down BFS, though there is no restriction from implementing a butterfly network with the newer bottom-up or direction optimizing BFS traversals as suggested by Beamer \emph{et al.} \cite{beamer2012direction}.




\section{Experimental Setup}
\label{sec:experiment-setup}

\begin{table*}[]
    \caption{Large graphs used in our experiments with execution times, GTEPS, and speedups. We report the average diameter for each graph. We do not include the smaller graphs used for testing other GPU frameworks in this table.}
    \tiny

    \begin{center}    
    \begin{tabular}{lrrrrrrrrrrrr}
    \textbf{Graph} & \multicolumn{1}{l}{\textbf{Vertices}} & \multicolumn{1}{l}{\textbf{Edges}} & \textbf{\begin{tabular}[c]{@{}r@{}}Ave. \\ Diam.\end{tabular}} & \multicolumn{1}{c}{\textbf{\begin{tabular}[c]{@{}c@{}}CPU \\ Time \\ (DO)\end{tabular}}} & \multicolumn{1}{c}{\textbf{\begin{tabular}[c]{@{}c@{}}CPU\\ GTEPs \\ (DO)\end{tabular}}} & \multicolumn{1}{c}{\textbf{\begin{tabular}[c]{@{}c@{}}CPU\\ Time \\ (TD)\end{tabular}}} & \multicolumn{1}{c}{\textbf{\begin{tabular}[c]{@{}c@{}}CPU\\ GTEPs \\ (TD)\end{tabular}}} & \multicolumn{1}{c}{\textbf{\begin{tabular}[c]{@{}c@{}}CPU-DO/\\ CPU-TD \\ Speedup\end{tabular}}} & \multicolumn{1}{c}{\textbf{\begin{tabular}[c]{@{}c@{}}DGX2\\ Time\end{tabular}}} & \multicolumn{1}{c}{\textbf{\begin{tabular}[c]{@{}c@{}}DGX-2\\ GTEPs\end{tabular}}} & \multicolumn{1}{c}{\textbf{\begin{tabular}[c]{@{}c@{}}DGX2-TD/\\ CPU-DO\\ Speedup\end{tabular}}} & \multicolumn{1}{c}{\textbf{\begin{tabular}[c]{@{}c@{}}DGX2-TD/\\ CPU-TD \\ Speedup\end{tabular}}} \\
    Webbase-2001   & 118.1M         & 1,019 M     & 375      & 0.43   & 2.37   & 0.48  & 2.11   & 1.13           & 0.18                  & 5.60  & 2.36           & 2.66 \\
    It-2004        & 41.2M          & 1,150.7M    & 26       & 0.31   & 3.72   & 0.33  & 3.49   & 1.07           & 0.03                  & 35.96 & 9.66           & 10.31     \\
    Uk-2005        & 39.4M          & 1,872.7M    & 21       & 0.25   & 7.49   & 0.33  & 5.67   & 1.32           & 0.03                  & 64.58 & 8.62           & 11.38     \\
    GAP\_twitter   & 61.5M          & 2,936.7M    & 14       & 0.22   & 13.41  & 1.55  & 1.89   & 7.08          & 0.09                  & 32.63 & 2.43           & 17.22     \\
    com-Friendster & 65.6M          & 3,612.1M    & 19       & 0.29   & 12.46  & 2.84  & 1.27   & 9.79          & 0.11                  & 32.84 & 2.64           & 25.82     \\
    GAP\_web       & 50.6M          & 3,860.5M    & 23       & 0.31   & 12.45  & 0.59  & 6.54   & 1.90           & 0.05                  & 84.01 & 6.75           & 12.84     \\
    GAP\_kron      & 134.2M         & 4,223.2M    & 5        & 0.29   & 14.56  & 3.04  & 1.39   & 10.48          & 0.01                  & 324.87                  & 22.31          & 233.85   \\
    GAP\_urand     & 134.2M         & 4,223.2M    & 7        & 0.47   & 8.99   & 3.20  & 1.32   & 6.81           & 0.20                  & 21.13 & 2.35           & 16.01     \\
    MOLIERE\_2016  & 30.2M          & 6,677.3M    & 15       & 0.61   & 10.95  & 4.31  & 1.55   & 7.07          & 0.19                  & 35.26 & 3.22           & 22.76    
    \end{tabular}

\end{center}
\label{tab:graphs}
\end{table*}

\begin{table*}[t]
    \caption{GPU and CPU system used in experiments.}
    \tiny
    \centering
    
    \begin{tabular}{|c|c|c|c|c|c|c|c|c|c|c|c|c|c|} \hline
    System & $\#$ GPUS & Processor & Micro & SM & SP  & Total  & DRAM & DRAM  &BW  & Form  & Power & Intercon. & Interconnect  \\ 
     &  & & arch. & & (per SM) & SPs  & Size & Type & (GB/s)  & Factor &  & &  BW (GB/s) \\ \hline \hline
    
    DGX2     & 16& V100  & Volta     & 80    & 64 & 5120 & 32GB    & HBM2 & 900 & SXM3 & 350 & NV Switch & 300\\  \hline
    \end{tabular}
    
    \vspace*{0.4 cm}
    
    \begin{tabular}{|c|c|c|c|c|c|c|c|c|c|} \hline
    System & Arch. & Micro & Processor & Frequency & Cores & Threads & LL-Cache & DRAM Size & DRAM Type    \\ 
     &  & arch. &  &  & & & &  &      \\ \hline \hline

    DGX2 & CPU x86-64       & Skylake   & $2 \times $ Intel Xeon Platinum 8168 & 2.7 GHz & $2 \times $ 24 & $2 \times 2 \times $ 24  & $2 \times $ 50 MB & 1536GB  & DDR4 \\  \hline
    \end{tabular}
    
    \label{tab:gpu-cpu-systems}
    \end{table*}

    \paragraph*{\bf{Volta GPUs}}
    
    The V100 is a Volta (micro-architecture) based GPU with 80 SMs and 64 SPs per SM, for a total of 5120 SPs (lightweight hardware threads). In practice, roughly 40K software threads are necessary for fully utilizing the GPU.
    The V100 has a total of 32GB of HBM2 memory and 6MB of shared cache between the SMs. Each SM also has a configurable shared memory of $96KB$. The V100 also has 640 tensor cores, though these are not used in our BFS algorithms. 
    The V100 has two form factors: PCI-E and SXM. PCI-E is the de facto form factor found in most consumer GPUs. The SXM form factor is equivalent to placing a GPU on a processor board, similar to a placing CPU on a motherboard. The SXM form factor has multiple NVLink channels allowing GPUs to communicate with other multiple GPUs concurrently at a higher than PCI-E. The SXM form factor GPU is known for outperforming its PCI-E counterpart due to increased frequency and power consumption.
    The V100 GPUs in our experiments are the SXM3 form factors and have a peak power consumption of 350 watts. Details of the GPUs can be found in \ref{tab:gpu-cpu-systems}.
    
    \paragraph*{\bf{DGX-2}}
    The NVIDIA DGX2 server is a single node server with sixteen Volta V100 GPUs. The DGX2 server was the first server to introduce NVSwitch. NVSwitch enables communication from each GPU to all the remaining GPUs for a total of 300GB/s of bandwidth. Each GPU has six incoming and outgoing links at 25GB/s (each). Thus, a GPU can send and receive 150GB/s concurrently. The fully connected network also ensures that the latency between the varying communication paths is uniform in length. The real benefit of NVSwitch over existing interconnects is that the on-device bandwidth and off-device bandwidth are within an order of magnitude of each other. This ensures that the communication within a DGX-2 is relatively balanced.
    Ang \emph{et al.} \cite{li2019evaluating} give a detailed performance analysis of NVSwitch and NVLink.
    The DGX-2 also has a high-end CPU processor, an Intel Xeon Platinum 8168 processor with 48 cores and 96 threads running at 2.7GHz. The DGX-2 used in our experiments has 1.5TB of DRAM memory.
    Additional details of the CPU can be found in \ref{tab:gpu-cpu-systems}.

    \paragraph*{\bf{Fanout }}
    Given the high-connectivity of the DGX-2 and its NVSwitch capability, we test our new ButterFly BFS algorithm with two different communication fanouts $f=1$ and $f=4$. A more thorough discussion of these fanouts is available in Sec. \ref{sec:sub_fanout}. 
    It is perhaps not surprising that the higher fanout counts givers better performance. The large fanout improves memory utilization while reducing the number of iterations.

    \paragraph*{\bf{Inputs}}
    Our tests use a mix of real-world and synthetic data taken from SuiteSparse \cite{davis2018suitesparse}. Table~\ref{tab:graphs} summarizes the properties of these graphs. All directed graphs get converted into undirected graphs, and we denote the total number of edges by $|E|$. In the conversion process, all duplicate edges and self-edges get removed. The number of edges after this process is denoted by $|\hat{E}|$. Our motivation for creating undirected graphs is that in undirected graphs, the largest connected component typically accounts for $90\%-95\%$ of the vertices and edges \cite{broder2000graph,    davis2018suitesparse, Snap-Stanford}. Thus, in most cases where the root is randomly selected, the traversed edges for that root are identical, which results in more consistent benchmarking.
    For each graph, we select 100 different random roots. We use the same roots across the different GPUS counts. We exclude the 25 fastest and 25 slowest times and report the average time for the remaining roots. Our analysis found that the execution times were particularly biased to the roots in the smaller connected components where these times could be less than $1\%$ the median execution time. To avoid this bias, we removed the fastest 25 execution times. When the root was in the largest connected component, we found that the execution times were fairly consistent, even for the {\bf slowest times}, which did not vary by more than 10$\%$ than the median execution time; thus our speedup analyis is fair.

    \paragraph*{\bf{Top Down Vs. Bottom-Up BFS}}
    It is a well-known fact that bottom-up BFS, with direction optimization, can be especially useful for a wide range of inputs, particularly random networks with power-law distribution. These optimizations are not always as beneficial for real-world networks, but they do offer potential speedups.
    In this paper, for the sake of simplicity, we focus on top-down BFS. Specifically, our initial goal was to check the butterfly communication network's viability and less on algorithmic optimization associated with coding. This also makes the benchmarking simpler as there are fewer control knobs.
    
    \paragraph*{{\bf GapBS \cite{beamer2015gap}}} The Graph Benchmarking Suite (GapBS for short) is a high performance collection of graph algorithms.
    GapBS is a highly tuned OpenMP framework designed for shared-memory CPU systems. It's implementations are some of the fastest known for the CPU.
    GapBS includes the direction optimizing BFS algorithm by Beamer \emph{et al.} \cite{beamer2012direction}. GapBS was recently evaluated by Azad \emph{et al.} \cite{azad2020evaluation} against several other fast frameworks including Galois \cite{pingali2011tao}, GraphIt \cite{zhang2018graphit}, and SuiteSparse \cite{davis2018suitesparse}. All the implementations in the study by Azad \emph{et al.} \cite{azad2020evaluation} are direction optimizing based, and GapBS is typically the fastest of these frameworks.  When GapBS is slower than the other frameworks, it is not slower by much. Therefore, comparing to GapBS ensures that we compare against the fastest shared-memory implementation on the CPU. In fact, the execution times we report in this paper are faster than the ones reported by Azad \emph{et al.} \cite{azad2020evaluation}.
    
    For benchmarking purposes, we added a top-down BFS implementation within GapBS. Thus, we can compare our new top-down ButterFly-BFS algorithm with both a top-down and direction-optimizing algorithm on the CPU. The average performance difference that we see in our CPU experiments between top-down and direction optimizing is similar to those reported in \cite{beamer2013distributed}. Further, the difference between BFS and betweenness centrality (which requires a top-down traversal) in the study by Azad \emph{et al.} \cite{azad2020evaluation} is roughly 20$\times$-30$\times$, therefore our benchmarking is similar to past report results.
      


    \paragraph*{\bf{Implementation Details}}
    
    Recall the discussion in \ref{sec:multi-gpu} where we compare two different approaches for multi-GPU algorithms. The OpenMP-based solution treats the compute node as a single shared memory system and the MPI solution treats each GPU as its own compute-node. Our algorithm is a hybrid of these two approaches.
    Our implementation is entirely in C++ and CUDA. Specifically, all parallel code on the GPU is in CUDA and parallelism on the CPU is OpenMP based. Each OpenMP thread is responsible for managing a different GPU. The CPU threads do not do any computational work. Instead, they are responsible for orchestrating the traversals on the GPU. Yet, each GPU's work is entirely done on the memory within that given GPU - similar to an MPI based solution.
    We selected OpenMP over MPI for CPU parallelism for several reasons. Our primary consideration was simplicity and programmability. Bulk data transfers between multiple GPUs using OpenMP can be done with calls to ``{\it cudaMemcpy}'' as each GPU can access all the other GPUs' memory. RDMAs and sends/receive messages are not necessary. Lastly, the different CPU threads can communicate with each other using shared-memory versus message passing\footnote{The programming model for managing the CPU is shared-memory and the programming model for the GPU is distributed}. All this makes the programming relatively simple and is one of the primary benefits of NVIDIA's DGX2 over a distributed system with an equal number of GPUs.
    While our solution targeted OpenMP, our algorithm is not limited to OpenMP and can be extended to MPI, which would enable multi-node execution. We will explore this in the future.

    \paragraph*{\bf{Graph Partitioning}}
    As a proof of concept, we use a straightforward 1D partitioning scheme where we divide the vertices to the multiple GPUs such that each GPU gets a near equal number of edges and the vertices are consecutive in their ids. For example, the vertex ids are  $\{0,1,2,...,i\},\{i+1,i+2,...,j\},\{j+1,j+2,..,k\},...$. The number of vertices on each of the GPUs can be quite different and will depend on the distribution of the adjacency arrays. Recall that in Alg. \ref{alg:pseudo-butterfly-bfs} each compute node checks if the traversed vertex belongs to its own set of vertices. Thus the algorithm can also work with 2D partitioning. For the sake of simplicity, both in concept and implementation, we focused on 1D partitioning.
    Furthermore, the difference in the number of vertices per compute node and the distribution on each compute node can lead to workload imbalance. For example, some iterations of BFS could require traversing a few small vertices on one GPU and many large vertices on a different GPU. This is a common problem in multi-node BFS traversals, though some techniques such as graph relabeling or partitioning can reduce their performance impact. In future work, we will investigate the benefit of graph partitioning and vertex relabeling. 
    
    In contrast to our inherently simple 1D partitioning, Gunrock \cite{wang2015gunrock} and Groute \cite{ben2017groute} use more advanced partitioning. Both frameworks use Metis \cite{karypis1995metis} for partitioning the graph to the multiple GPUs. Metis is entirely CPU-based, and as such, the partitioning is executing using the CPU connected to the GPU. For both Groute and Gunrock, we do not include the graph partitioning execution time when compare the BFS traversal. The partitioning cost varies and is equivalent to running several thousand BFS traversals from different roots using our new algorithm.

    \paragraph*{\bf{Hornet}}
    \label{sec:hornet}
    The Hornet data-structure \cite{green-hornet} was first introduced as a framework for solving dynamic graph problems. Though the original framework also worked well for static graph problems, such as BFS. Hornet offers a high-level API that makes accelerated graph algorithms reasonably easy to implement. Through this API, which consists of parallel primitives, it is possible to implement a high performing BFS algorithm in a matter of twenty lines of code \footnote{Not including variable instantiation and memory allocation.}. 
    These primitives hide many complicated parallel programming nuances and allow algorithm designers to implement their algorithms by mapping their problems to a small number of templated functors.
    
    Hornet currently supports a static graph back-end, similar to CSR; this back-end removes the need to create a more complex dynamic graph data structure. Algorithms in Hornet can work for both the static and dynamic back-ends by changing a single flag in the initialization process.

    Our multi-GPU ButterFly BFS algorithm is implemented using the Hornet framework. We report performance numbers for the static back-end, though we found little difference in performance between the static and dynamic back-ends in the algorithm. We found that the ETL process for the static back-end is more efficient than the dynamic back-end.
    
    \paragraph*{{\bf Load Balanced Traversals Per compute-node}}
    
    The BFS traversal on GPU uses the recent Logarithmic Radix Binning (LRB) technique discussed in \cite{green-lrb-2019,green-tri-lrb}. LRB groups vertices in the frontier into roughly 32 bins or 64 bins, depending on the graphs' maximal number of vertices. Vertices in the same bin have an adjacency list that is never more than twice as big or small as any other vertices in that bin. Using LRB, we can dispatch multiple GPU kernels using multiple GPU streams, one for each bin, where the number of threads in the thread block is decided based on the upper bound of the adjacency array for that bin. 

\section{Empirical Performance Analysis}
\label{sec:experiments}

\begin{figure*}[t]
    \centering
    
    \subfloat[webbase-2001]{\includegraphics[width=0.33\columnwidth]{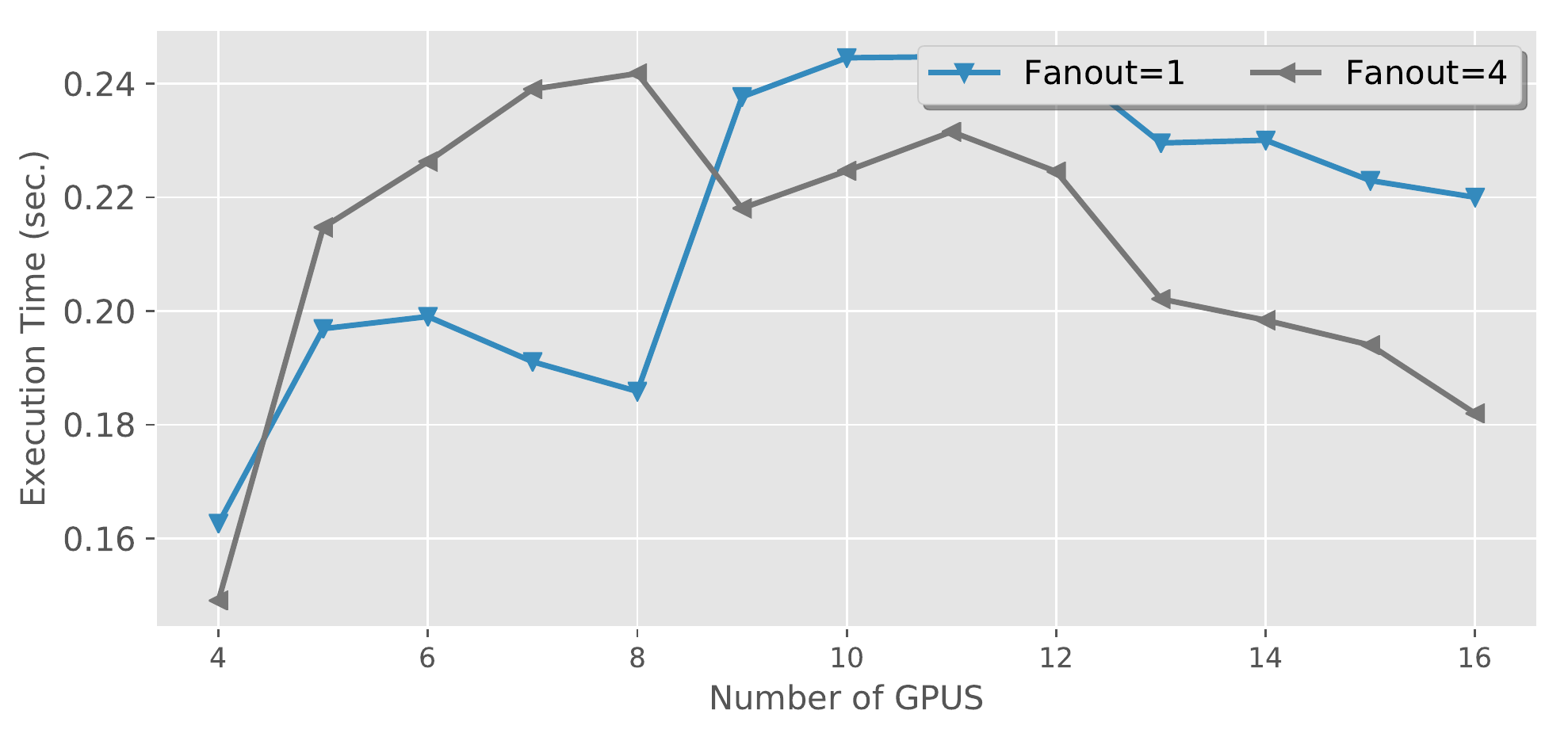}}
    \subfloat[it-2004]{\includegraphics[width=0.33\columnwidth]{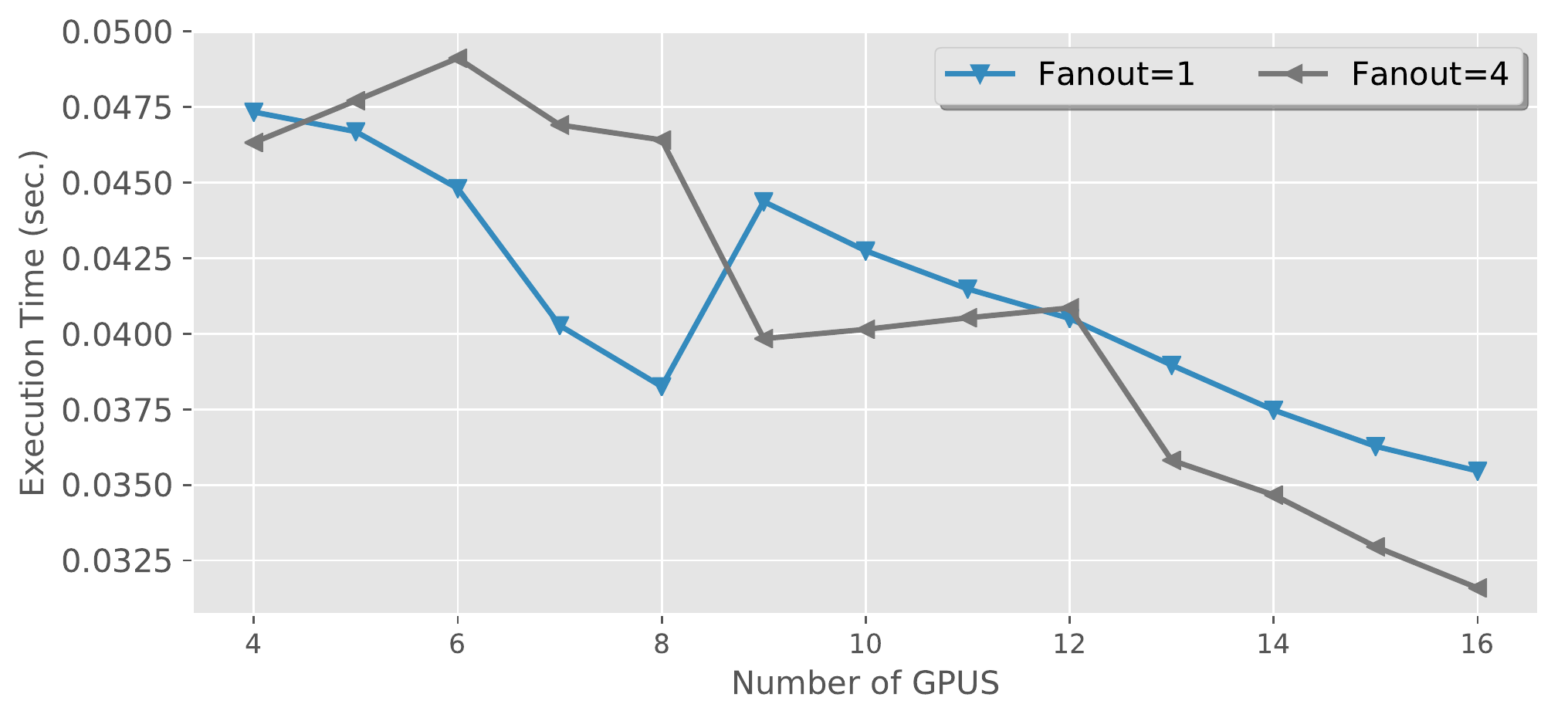}}
    \subfloat[uk-2005]{\includegraphics[width=0.33\columnwidth]{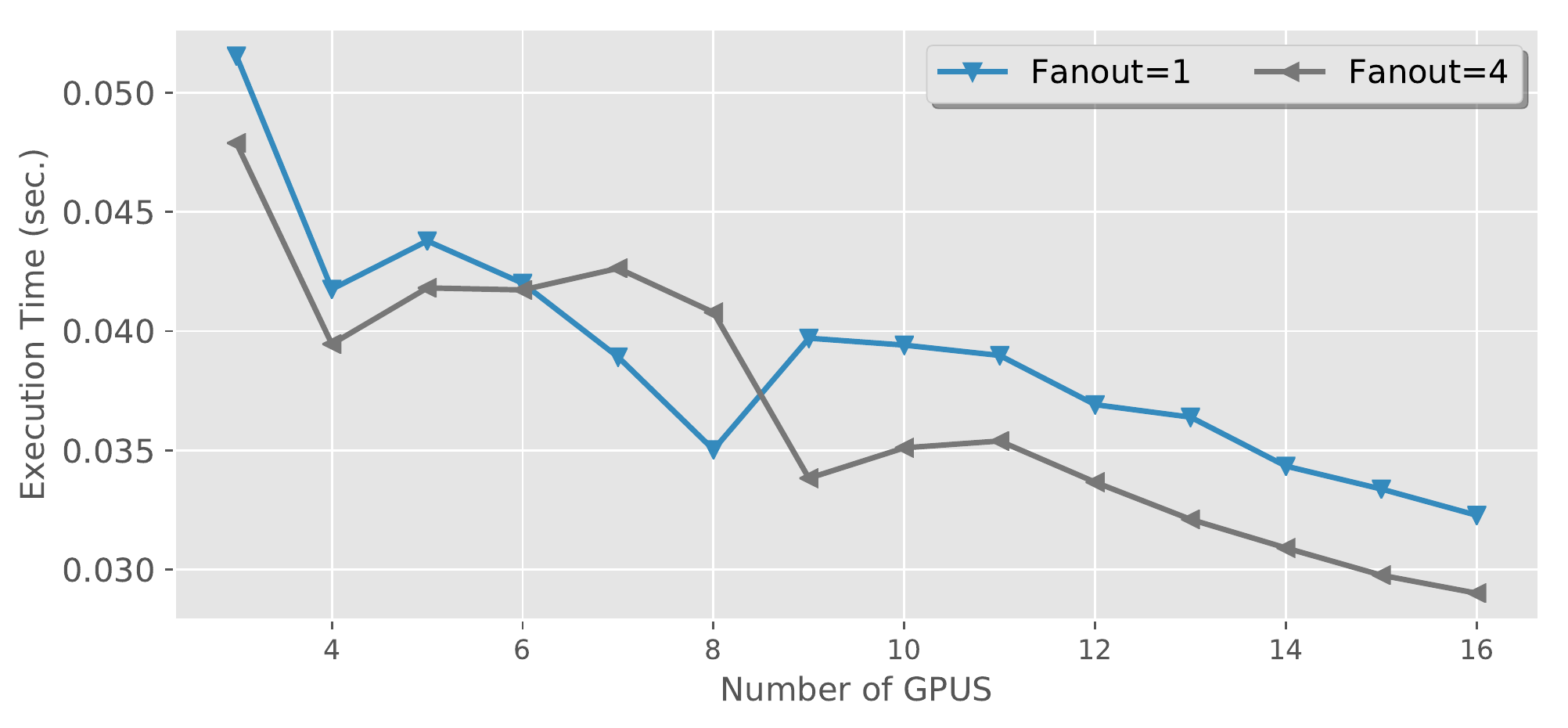}}
    
    \subfloat[GAP-twitter]{\includegraphics[width=0.33\columnwidth]{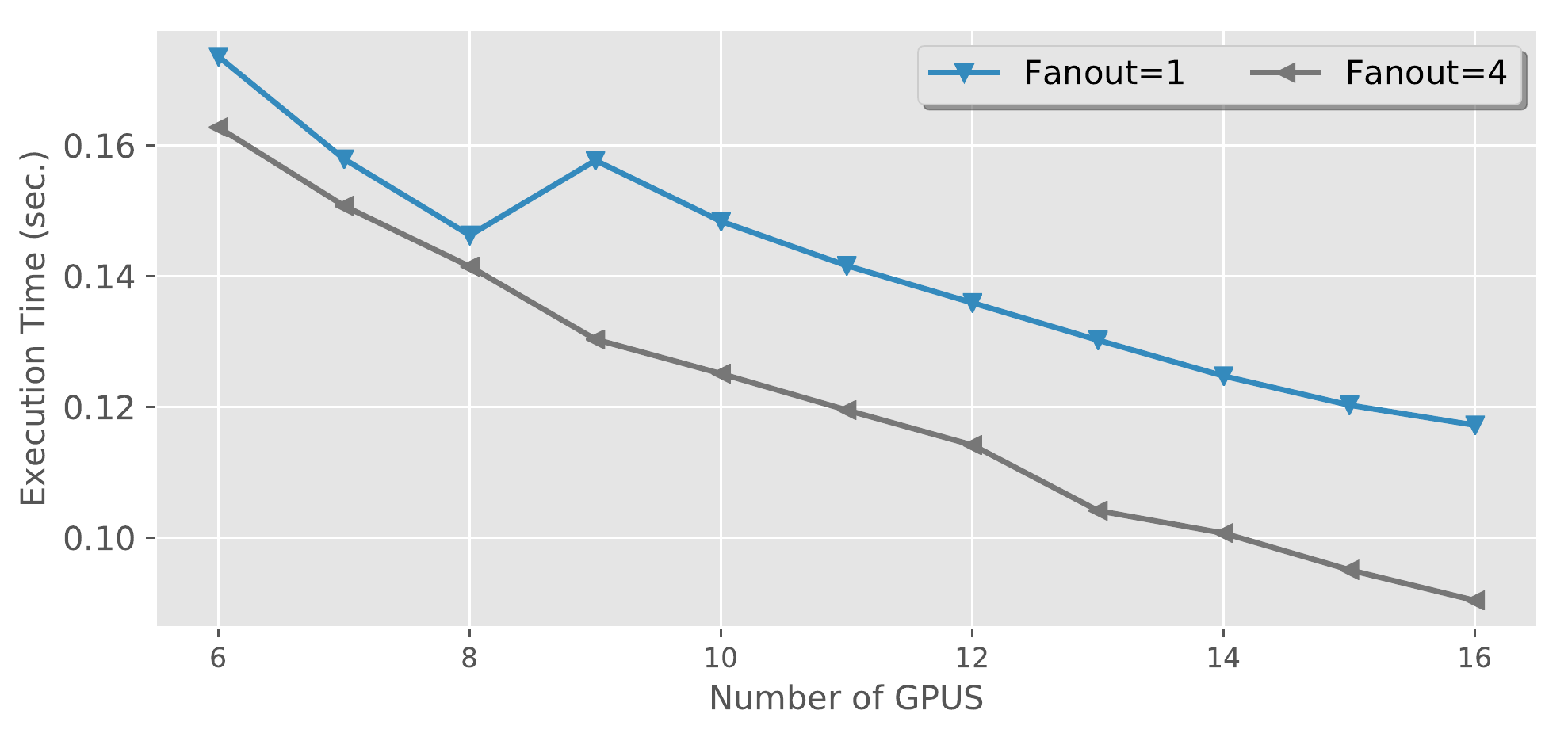}}
    \subfloat[com-Friendster]{\includegraphics[width=0.33\columnwidth]{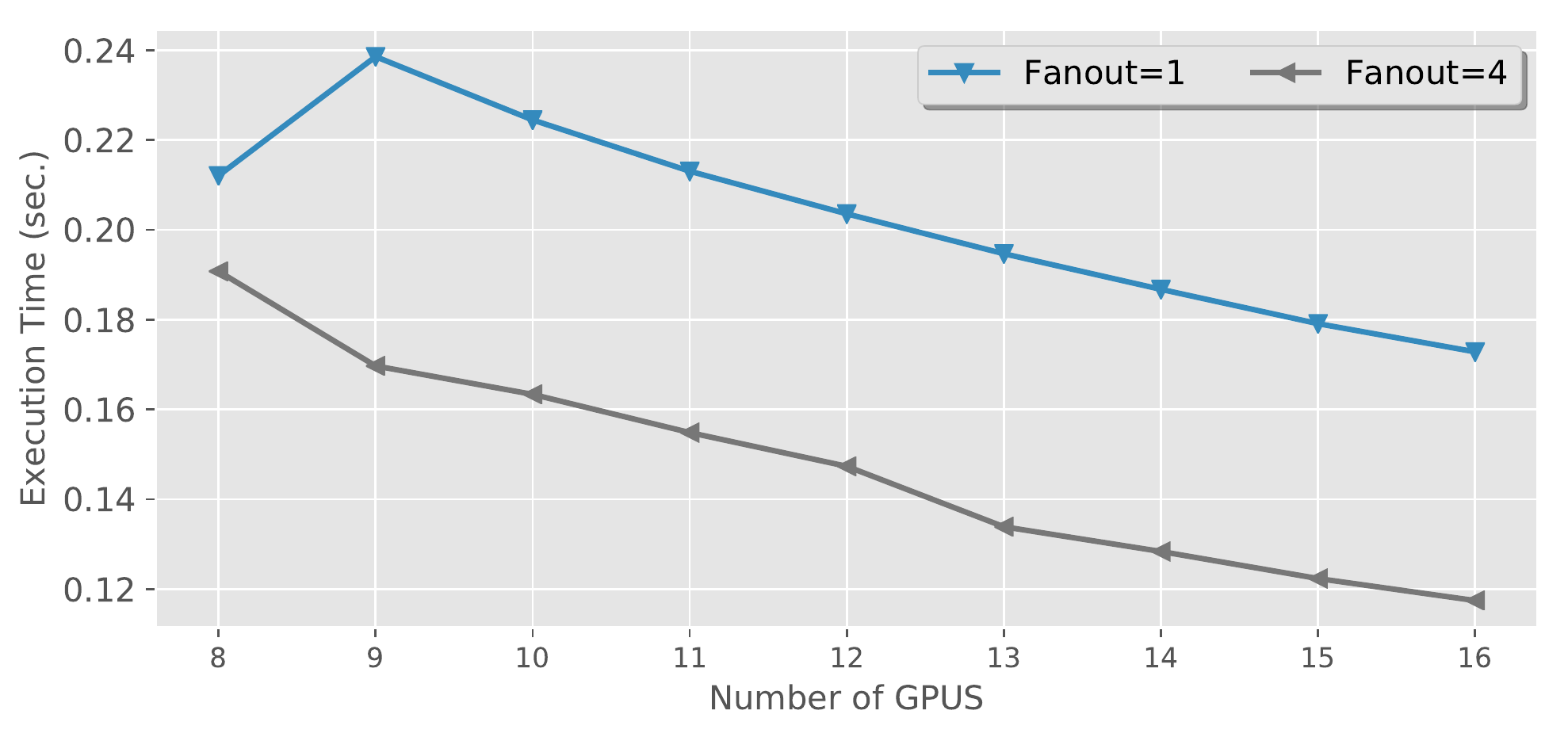}}
    \subfloat[GAP-web]{\includegraphics[width=0.33\columnwidth]{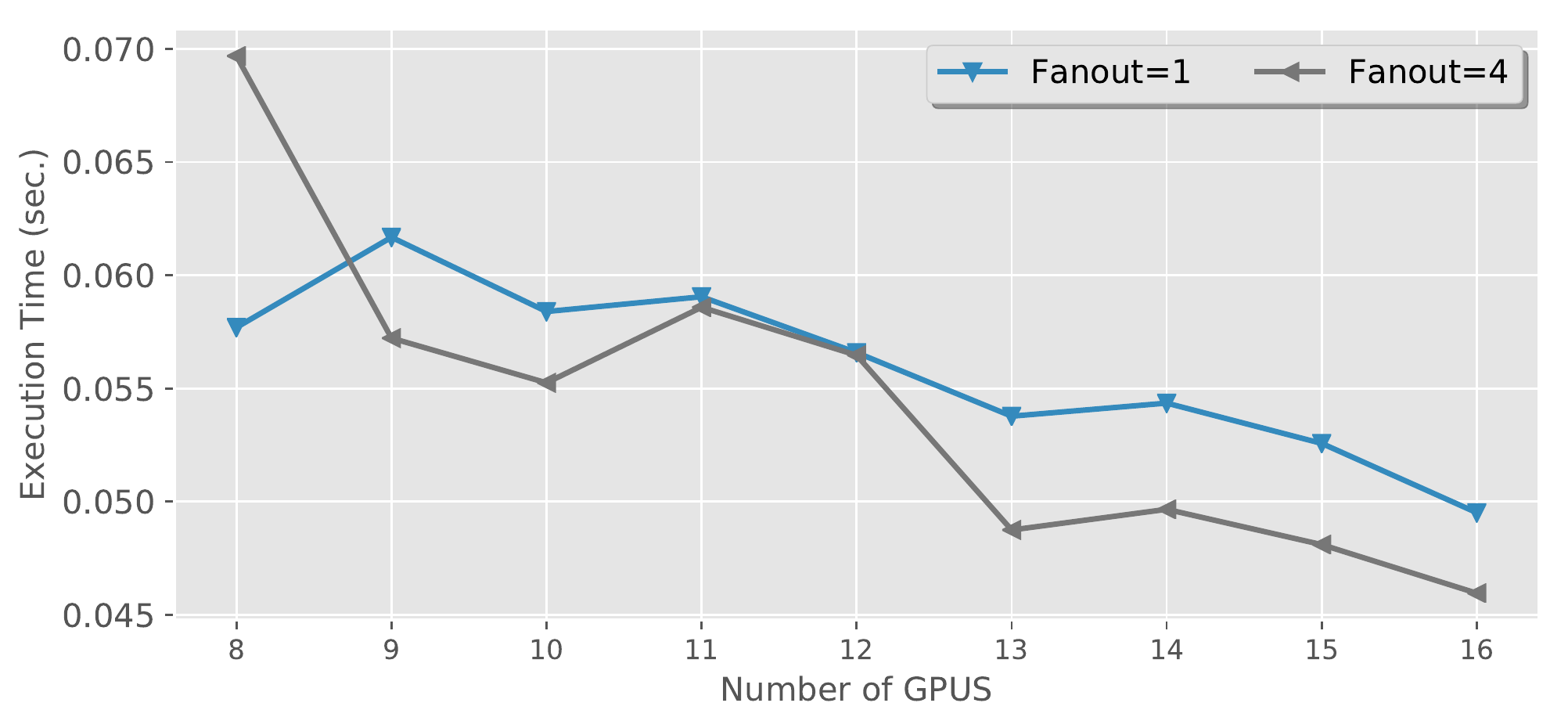}}
    
    \subfloat[GAP-kron]{\includegraphics[width=0.33\columnwidth]{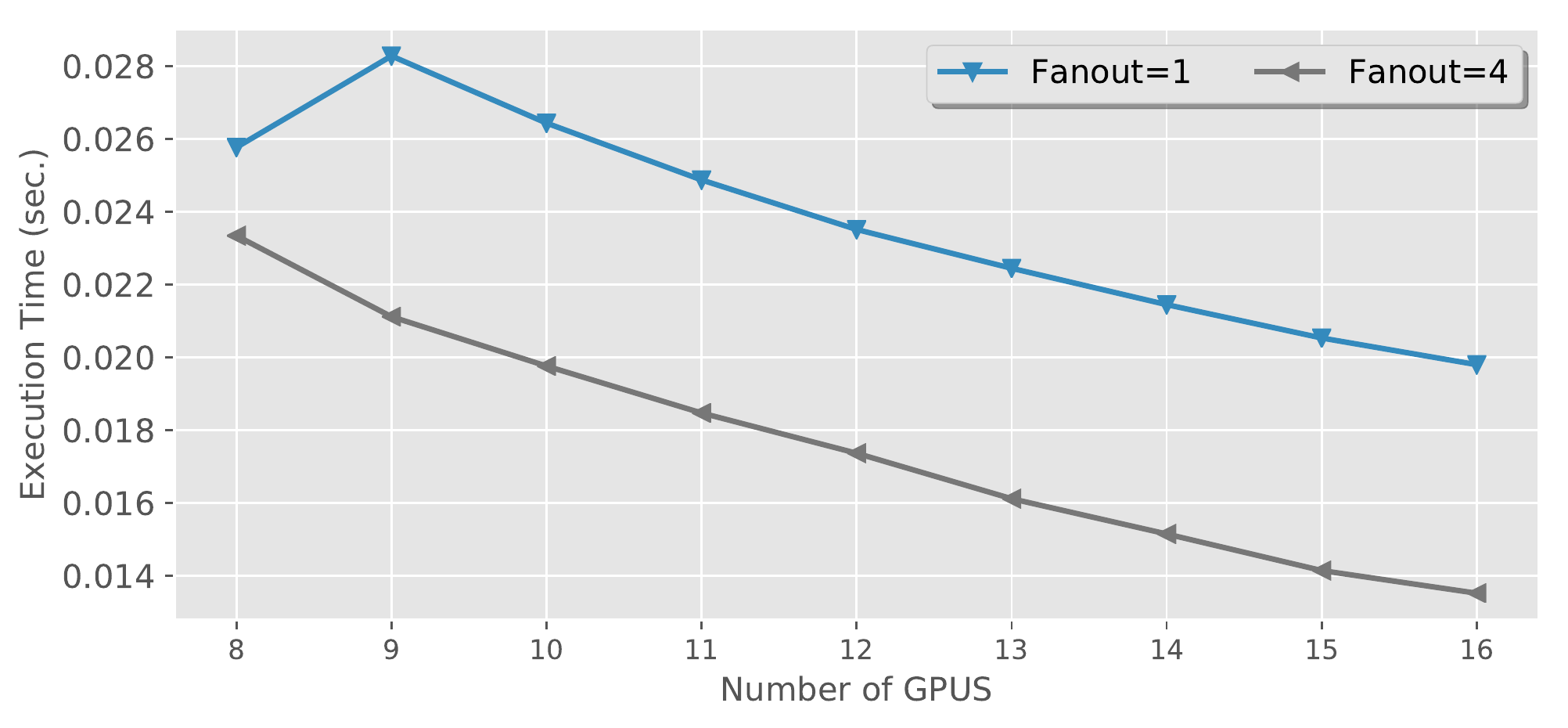}}
    \subfloat[GAP-urand]{\includegraphics[width=0.33\columnwidth]{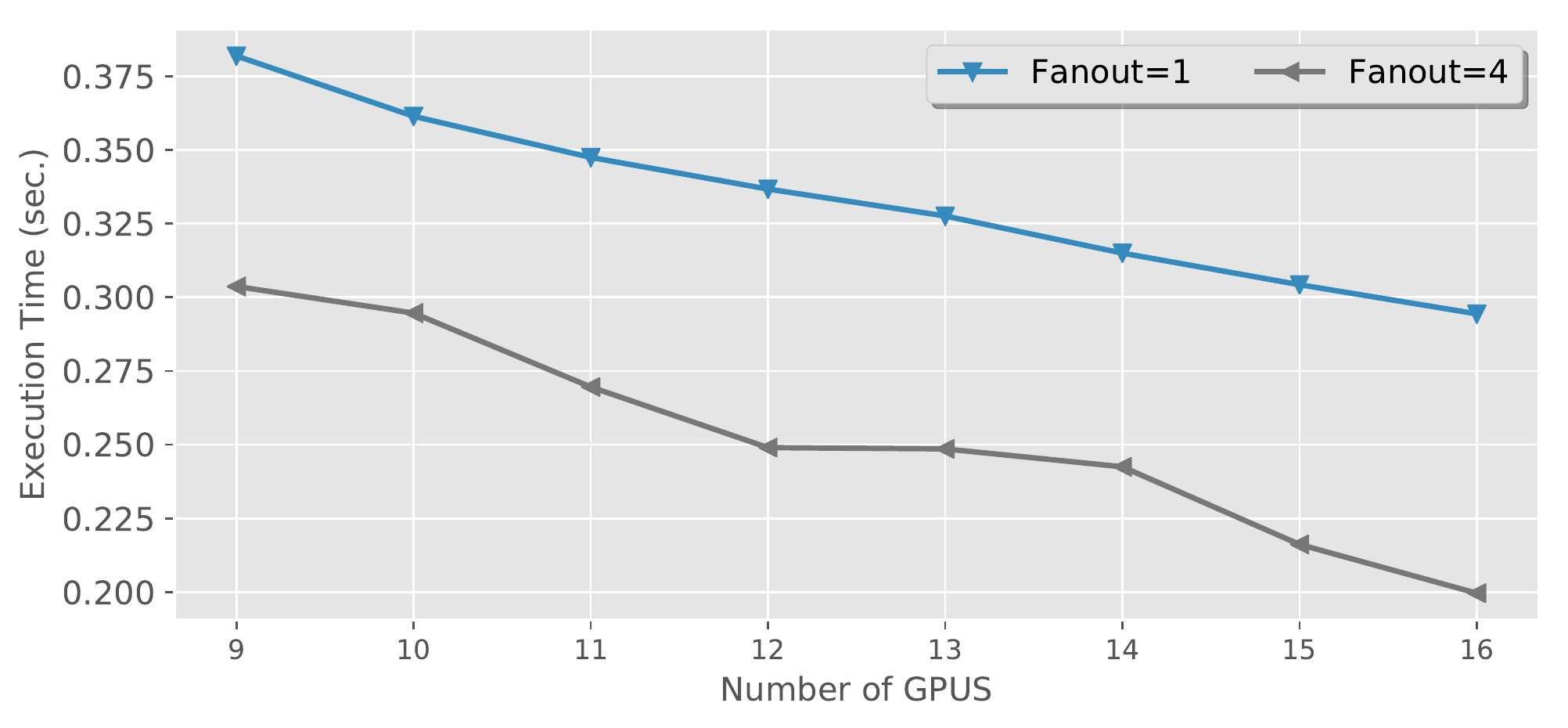}}
    \subfloat[moliere-2016]{\includegraphics[width=0.33\columnwidth]{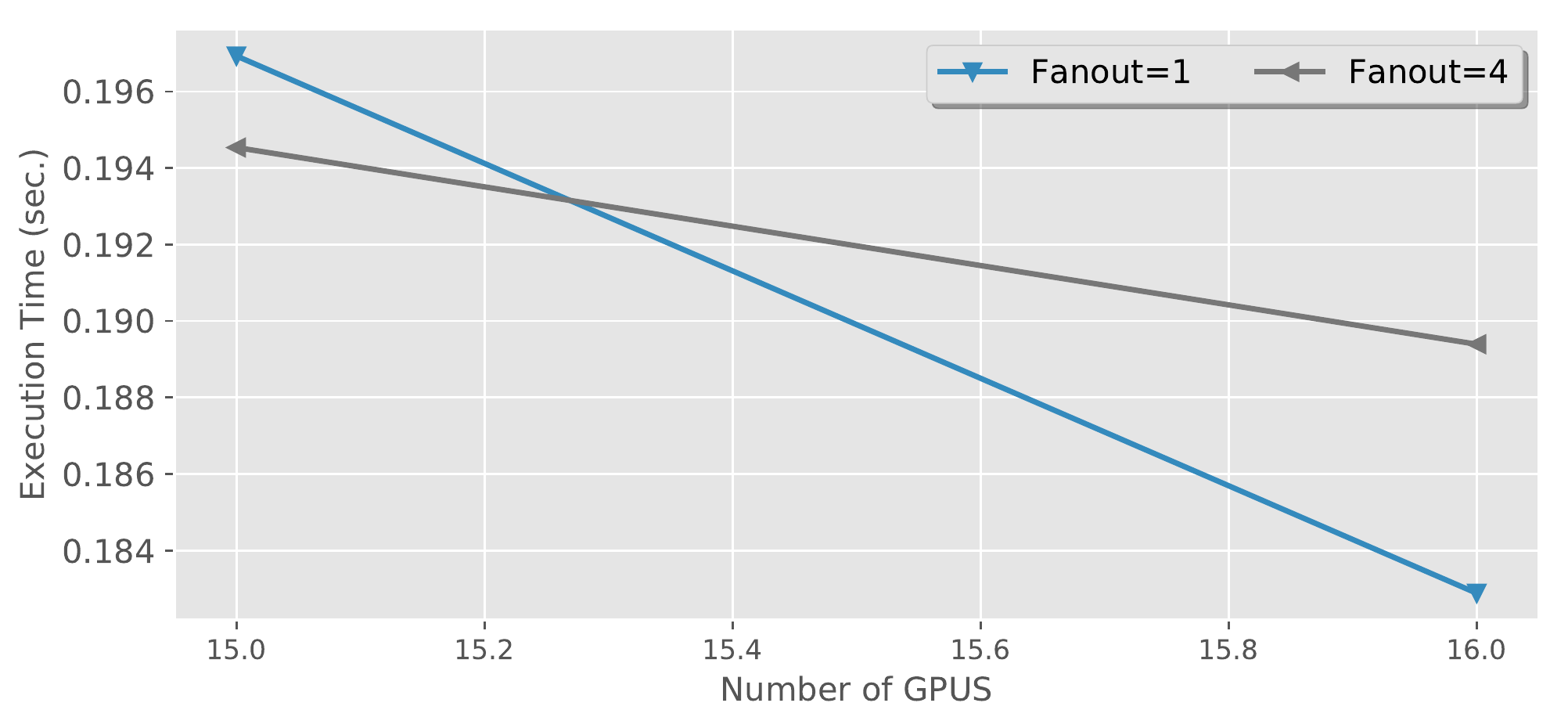}}
    
    \caption{Execution time of ButterFly BFS on a DGX-2 with 16 GPUs. Plots are organized from graphs with the least number of edges to the graph with the largest number of edges.}
    \label{fig:time-bfs}
    \end{figure*}

    \paragraph*{{\bf DGX-2 Performance Comparison with a high-end CPU}}
    We start by comparing our new ButterFly-BFS algorithm running with a single DGX-2 server with a high-end Intel x86 CPU performance. 
    Both of these systems are shared-memory systems. Previous BFS algorithms were limited to the memory size of a single GPU. Thus, CPU shared-memory systems could process significantly larger graphs. ButterFly-BFS now enables graphs of roughly the same within a shared-memory architecture.
    The x86 server used in our experiments has 48 cores (96 threads) with 1.5 TB of DRAM. 
    The DGX-2 server has 16 GPUs, each with 32GB of HBM2 memory for 512GB of HBM2-DRAM. Thus, from a memory perspective, these two systems can process graphs of roughly the same size.

    For GapBS, we benchmarked two different configurations: classic top-down BFS and direction optimizing BFS. For GapBS, we report execution with all cores and threads utilized in the execution. Table \ref{tab:graphs} depicts the execution time for both these configurations. The performance of direction optimizing BFS is highly dependent on several internal parameters that decide when the algorithm swaps between top-down and bottom-up. We use the default parameters, but note that manually tuning these parameters per graph can improve performance. Typically this is undesirable as finding the ideal set of parameters requires multiple runs of BFS algorithms.  
    Even the default values increase the performance of direction optimizing BFS over the classic top-down algorithm, depicted in the column with the speedup of direction-optimizing (DO) over top-down (TD). In several cases, the speedup was over $20\times$ and was as high as $86\times$ (for a random graph). In other cases, the speedup was only $10\%-90\%$ greater than top-down (which is still very desirable). Direction optimizing BFS loses a lot of its benefit in large diameter graphs. Specifically, these graphs tend to have limited parallelism, and there is no way to avoid traversing some edges.
    The key takeaway message is that direction optimizing BFS can offer performance improvement. 
    
    The two rightmost columns in Table \ref{tab:graphs} depict the speedup of our new ButterFly BFS algorithm using a classic top-down traversal on a DGX-2 server in comparison to the GapBS implementation for both direction-optimizing (before the last column) and top-down (last column).
    ButterFly BFS outperforms direction optimizing BFS on the CPU anywhere from $2\times - 22\times$. For most cases where the speedup was only $2\times - 3\times$ faster, we can see that the direction optimization on the CPU added a speedup of over $5\times$ compared to top-down.

    ButterFly BFS outperforms top-down BFS on the CPU anywhere from $2\times - 233\times$. For most inputs, the performance increase was in the range of $10\times - 25\times$.
    Comparing the two top-down algorithms highlights the scalability and performance benefit of ButterFly BFS as both algorithms use the same traversal methodology, and both implementations are designed for single-node systems.  
    
    The graph {\it Webbase-2001} is especially interesting as the traversal rates for both the CPU and DGX-2 were low. {\it Webbase-2001} has a reasonably large diameter; to be exact, it has a large tail of about one hundred vertices long - one at each level. Thus, there is no available parallelism, and the synchronizations dominate the execution time. 
    In contrast, the {\it GAP$\_$kron} network has is small diameter graph; therefore, there are fewer synchronization overheads and few iterations.     
    
    
    \paragraph*{{\bf Strong Scaling Performance Analysis}}
    
    We explore the scalability of our new ButterFly BFS as a function of GPUs. We report execution numbers for both a fanout of one and fanout of four, depicted in Fig. \ref{fig:time-bfs}. For each graph, we report execution times for the smallest number of GPUs that could be used for the traversal. In some cases, the graphs could fit into the memory of a smaller number of GPUs; however, the ETL process typically increased the memory footprint by $2\times-3\times$. Our goal was not to optimize the ETL pipeline --- which is a challenge in its own right. As a rule of thumb, each GPU was responsible for 500M edges. Therefore, the minimal number of GPUs needed is the number of edges in the graph divided by 500M.
    
    In the first row of the plot are the smallest networks used in our experiments. All three of these networks have 1B-2B edges. Of these networks, both {\it it-2004} and {\it uk-2005} show some performance improvements as more GPUs are added, though the performance gains are not high. The limited performance increase is partially due to the small amount of work per GPU for the higher GPUs counts ($60M-120M$ edges). {\it Webbase-2001}'s high diameter makes it especially tough to parallel as the GPUs are mostly idle.
    There is a clear and steady performance improvement for the remaining six networks as the number of GPUs increases, especially for a fan out of four.

    \paragraph*{{\bf Fanout Difference}} For most graphs and GPU counts, the fanout four outperforms the fanout one as the communication phase of the algorithm is more efficient. For several networks, using a large fanout increases the total performance by over $50\%$  for larger GPU counts. In particular, the increase in the number of GPUs also increases the number of communication rounds needed for synchronizing the frontiers. For 16 GPUS, four rounds of communication are necessary for a fanout one compared to two rounds of communication required by fanout four.  
    Furthermore, note that there is typically a performance loss for fanout one going from eight GPUs to nine GPUs. The performance loss is attributed to a communication bottleneck in the last round of the communication where eight GPUs need to communicate with the remaining GPU (Fig.\ref{fig:fanout1} (f)). This bottleneck does not happen for the larger fanout four with 16 GPUS. 
    For the {\it moliere-2016} network, fanout one slightly outperforms fanout four. We ran this test multiple times to confirm this result. It could be that the vertex queuing in the butterfly network is the bottleneck.  While the difference is not significant, it does require additional investigation. 
    
    \paragraph*{\bf{Speedup Analysis}}
    
    We start by noting that there is a clear performance trend for the bigger graphs that as the number of GPUs increases, the execution times reduce. While it is evident that this trend is desirable, many distributed BFS algorithms do not have this property. ButterFly BFS does achieve perfect linear speedups, yet it shows improved performance as more nodes are added; this is in stark contrast to both the algorithms in Gunrock \cite{wang2017gunrock} and Groute \cite{ben2017groute} where the execution increases with the number of GPUs. Both Gunrock and Groute need to use dynamic memory allocations for the buffers used for transferring the frontiers. In contrast, ButterFly BFS place's an upper bound on the memory requirements based on the fanout. 

    As most of the graphs were too large to fit into a single GPU's memory, we could not measure the speedup compared to a single GPU. 
    Instead we measure the relative speedup out of the ideal speedup as follows. We measure the relative speedup for each network as follows: $Speedup = t_{min} \div t_{max}$, where
    $t_{max}$ is the execution with the maximal number of GPUs (16 in our case) and $t_{min}$ is the execution for the minimal of GPUs used for a given network. The ideal speedup is given by: $Ideal= max\_nodes \div min\_nodes$.
    Lastly, we define the speedup utilization as $Utilization = Speedup \div Ideal$.
    
    For each of the larger graphs we report the following numbers for fanout four: ($Speedup$, $Ideal$, $Utilization$):
    {\it GAP-Twitter} (1.8, 2.66, $67.5\%$), 
    {\it com-Friendster} (1.62, 2, $81.2\%$), 
    {\it GAP-web} (1.53, 2, $76.6\%$), 
    {\it GAP-kron} (1.77, 2.0, $88.4\%$), 
    {\it GAP-urand} (1.518, 1.7771, $85.4\%$), and
    {\it moliere-2016} (1.025, 1.06, $96.2\%$).
    Altogether, ButterFly BFS achieves good speedups and good system utilization.

    \paragraph*{{\bf Other Multi-GPU BFS Algorithms}} 
    As part of our benchmarking, we used several existing multi-GPU BFS algorithms that are known to have good performance, including the algorithm by Pan \emph{et al.} \cite{pan2017multi} which was part of an earlier version of Gunrock \cite{wang2017gunrock} and the algorithm in Groute \cite{ben2017groute}. Both Gunrock and Groute, are designed for multiple-GPUs within a single compute node. Both use OpenMP for managing the multiple threads. In contrast, the algorithms by Hang \& Howie \cite{liu2015enterprise} and Bisson \emph{et al.} \cite{bisson2015parallel} are designed for distributed systems and are MPI based. All these algorithms use direction optimizing traversals. However, going from one GPU to multiple GPUs, these implementations no longer benefit from the faster traversal as the communication becomes the execution bottleneck.
    
    We were only able to test Gunrock and Groute for smaller graphs than those presented in Table \ref{tab:graphs}. We ran into multiple execution problems for the large graphs. The algorithms by Pan \emph{et al.} \cite{pan2017multi} and Bisson \emph{et al.} \cite{bisson2015parallel} were unable to load large graphs without their ETL process failing. The algorithm by Hang \& Howie \cite{liu2015enterprise} uses a nonstandard file format, and we were unable to convert the 50GB files to the required format. Both Gunrock and Groute use Metis for partitioning the graph and Metis runs into allocation problems with graphs with 2B edges.
    
    We tested Gunrock and Groute using the ``kron\_g500-logn21'' network (two million vertices and 182 million edges). Similar to the results report by Ben-Nun \cite{ben2017groute}, we found that Groute was faster than Gunrock. Our new algorithm was about $3\times$ faster than Groute for 6 GPUs. Groute did not finish for larger GPU counts. In contrast, Gunrock finished with 16 GPUs; however, its execution time increased with each additional GPU. Our algorithm using 16 GPUs was over $50\times$ faster than Gunrock using an equal number of GPUs.  
    
    We were able to test the algorithm by Bisson \emph{et al.} \cite{bisson2015parallel} using random graphs; however, these graphs are not identical to the graphs taken from the SuiteSparse collection. As such, we are not able to give an accurate apples-to-apples comparison. If one merely evaluates the traversal rate, we found that our new algorithm was faster than Bisson \emph{et al.} \cite{bisson2015parallel}.

    
    

    
    


\section{Summary}

In this paper we introduced a novel communication pattern, namely a butterfly network, for synchronizing the frontiers of a BFS traversal acros multiple compute-nodes. Butterfly BFS offers a new approach for reducing the number of messages sent, the sizes of the messages, and reduces the buffer sizes needed in the communication. In fact, 
Butterfly BFS offers a tight bound on the memory consumption. Butterfly BFS greatly reduces the network saturation by sending fewer messages.

We showed the buttefly network can be implemented with multiple fanouts that enable utilizing multiple network adapters and communication lines (if and when available). We also showed this emperically that with a fanout of four, performance could increase by as much as $50\%$ while removing performance bottlenecks in comparison to a fanout of one.

In our emperical analysis we show that the performance of Butterfly-BFS continues to improve as more compute nodes are are added ($75\%$ utilization of the peak scalability).
This is a non trivial feature for many multi-node BFS algorithms. While our implementation of Butterfly-BFS uses a top-down BFS traversal, it outperforms 
GapBS \cite{beamer2015gap,beamer2012direction}, one of the fastest known BFS implementations for shared-memory systems which uses a direction-optimizing traversal, anywhere from $2\times - 22\times$ faster and for most cases it is over $6 \times$ faster. When we compare Butterfly-BFS with GapBS using a top-down traversal, our algorithm is $2\times - 233\times$ faster, with a typical speedup in the range of $10\times - 25\times$. Both Butterfly-BFS and GapBS utilize a single server, though our Butterfly-BFS uses sixteen GPUs found in the DGX-2 server. 
Lastly, we showed that our Butterfly-BFS can achieve a traversal rate of 300 GTEP/s for scale 29 Kronecker network. 

Altogether, Butterfly-BFS offers a new and promising traversal mechanism for multi-node traversals. In this work, we showed performance for an OpenMP and CUDA implementation as all our GPUs we co-located within the same server. While we greatly benefited from NVIDIA's NVSwitch interconnect for communication across the GPUs. Each GPU was responsible for managing its own buffer and communication, similar to how CPUs and GPUs behave in a fully distributed environment. Thus, Butterfly-BFS is not limited to single-node servers and can be extended for fully distributed systems.

\bibliographystyle{ACM-Reference-Format}
\bibliography{bibfile,green,bader,bfs}




\end{document}